\documentclass[%
reprint,
superscriptaddress,
amsmath,amssymb,
aps,
pra,
floatfix,
]{revtex4-2}

\usepackage{graphicx}
\graphicspath{{images/}}
\usepackage{xcolor}
\usepackage[normalem]{ulem}
\usepackage{dcolumn}
\usepackage{bm}
\usepackage{mathtools}
\usepackage[thinc]{esdiff}
\usepackage{textcomp}
\usepackage{bbold}
\usepackage{braket}
\usepackage{hyperref}
\usepackage{dsfont}
\usepackage{amsthm}

\usepackage{placeins}
\usepackage{siunitx}

\definecolor{LB}{RGB}{66,10,146}    

\begin{document}


\title{Ion Trap Long-Range XY Model for Quantum State Transfer and Optimal Spatial Search}

\author{Dylan Lewis}
 \email{dylan.lewis.19@ucl.ac.uk}
\affiliation{
 Department of Physics and Astronomy, University College London, 
 London WC1E 6BT, United Kingdom
}

\author{Leonardo Banchi}
\affiliation{Department of Physics and Astronomy, University of Florence, 
via G. Sansone 1, I-50019 Sesto Fiorentino (FI), Italy}
\affiliation{INFN Sezione di Firenze,  via G. Sansone 1,  I-50019 Sesto Fiorentino (FI), Italy}

\author{Yi Hong Teoh}
\affiliation{Institute for Quantum Computing and Department of Physics and Astronomy, University of Waterloo, Waterloo, Ontario N2L 3G1, Canada}

\author{Rajibul Islam}
\affiliation{Institute for Quantum Computing and Department of Physics and Astronomy, University of Waterloo, Waterloo, Ontario N2L 3G1, Canada}

\author{Sougato Bose}
\affiliation{
 Department of Physics and Astronomy, University College London, 
 London WC1E 6BT, United Kingdom
}


\begin{abstract}
Linear ion trap chains are a promising platform for quantum computation and simulation. The XY model with long-range interactions can be implemented with a single side-band Mølmer-Sørensen scheme, giving interactions that decay as $1/r^\alpha$, where $\alpha$ parameterises the interaction range. Lower $\alpha$ leads to longer range interactions, allowing faster long-range gate operations for quantum computing. However, decreasing $\alpha$ causes an increased generation of coherent phonons and appears to dephase the effective XY interaction model. We characterise and show how to correct for this effect completely, allowing lower $\alpha$ interactions to be coherently implemented. Ion trap chains are thus shown to be a viable platform for spatial quantum search in optimal $O(\sqrt{N})$ time, for $N$ ions. Finally, we introduce a $O(\sqrt{N})$ quantum state transfer protocol, with a qubit encoding that maintains a high fidelity. 
\end{abstract}
\maketitle

\section{\label{sec:intro}Introduction}
Quantum information processing requires qubits that can be coherently and precisely controlled and measured~\cite{divincenzo_physical_2000}. Linear chains of atomic ions that are trapped by electromagnetic fields and held in vacuum chambers can fulfil these requirements and have been established as an exciting and promising platform for quantum computing~\cite{steane_ion_1997,monroe_scaling_2013, brown_co-designing_2016}. The qubit can be encoded in hyperfine or Zeeman ground states with the ions experiencing a spin-dependent force via the Mølmer-Sørensen scheme~\cite{sorensen_entanglement_2000}. Virtual phonons then mediate spin-spin interactions between the ions due to the Coulomb force~\cite{blatt_entangled_2008}. In this way the ion-trap chains become natural platforms for the quantum simulation of spin-spin interacting systems~\cite{monroe_programmable_2021}. Significant research interest has focused on engineering specific Hamiltonians for quantum simulation~\cite{porras_effective_2004,kim_quantum_2011,blatt_quantum_2012,wall_boson-mediated_2017, kiely_relationship_2017}. 
Particularly unique are XY spin models with long-range interactions that decay as $1/r^\alpha$, where $\alpha$ is a tunable parameter. This model suffers from coherent leakage outside the model space, particularly for small $\alpha$. Here, we show how to fully mitigate for this coherent error and provide two applications: optimal spatial quantum search, and a $O(\sqrt{N})$ quantum state transfer protocol.

Optimal spatial search is the problem of finding a particular marked node on a graph in optimal $O(\sqrt{N})$ time for $N$ nodes. The graph can be encoded physically with the locations of single excitations as the nodes of the graph and the edges of the graph describing the possible hopping of the single excitation -- equivalent to the XY model in the single-excitation subspace. Childs and Goldstone~\cite{childs_spatial_2004} found that the spatial search problem for the complete graph, the hypercube graph, and $d$-dimensional periodic lattices of $d>4$ can be solved in optimal time using continuous-time quantum walks. A number of high dimensional graphs were subsequently found that permit optimal spatial search~\cite{childs_spatial_2014,novo_systematic_2015,chakraborty_spatial_2016,chakraborty_optimal_2017,novo_environment-assisted_2018,wong_quantum_2018,osada_continuous-time_2020,sato_scaling_2020, chakraborty_optimality_2020, malmi_spatial_2022}. Recently, optimal spatial search in one dimension using long-range interactions was found to be possible~\cite{lewis_optimal_2021}. 
However, does this translate to a physical implementation? Here, we answer this affirmatively by proposing the experimental details for ion-trap chains to realise optimal spatial search and a related scheme for state transfer. 

The quantum state transfer protocol we introduce is described in detail in Section~\ref{sec:protocol}. The protocol provides transfer in $O(\sqrt{N})$ time with unit fidelity asymptotically for $\alpha <1$ and with decreasing fidelity for $1<\alpha < 1.5$, as opposed to $O(N)$ for state transfer on nearest neighbour interacting chains. A relevant question for a quantum bus is how fast can quantum communication be performed along a spin chain in the single-excitation subspace with long range interactions? The fastest possible communication protocol must be bounded by the speed at which correlations can spread. Recently, the Lieb-Robinson bounds for long-range interacting systems with $\alpha > d$, where $d$ is dimension, have been established~\cite{foss-feig_nearly-linear_2014,else_improved_2018,tran_locality_2018, chen_finite_2019, kuwahara_strictly_2020, tran_lieb-robinson_2021}. The bounds for the interacting distance against time give an effective light-cone of interactions. For spin chains in $d=1$, these results characterise the light-cone for $\alpha > 1$. State transfer cannot occur in a time faster than the scaling limit imposed by these bounds. Fast state transfer protocols have been found that saturate these bounds for all $\alpha > 1$~\cite{tran_hierarchy_2020, tran_optimal_2020}. In the free-fermion case, where particles are non-interacting, reduced bounds have also recently been established~\cite{tran_hierarchy_2020}. In this case, as $\alpha \rightarrow 1.5$ the minimum time for correlations to spread approaches $t \sim \sqrt{N}$ – the same scaling as our protocol. For $\alpha < 1.5$, we find that our protocol does not saturate the bound. However, our protocol is notably simpler experimentally, being a \emph{time-independent} Hamiltonian, and we show in detail how it can be implemented. The reduction in control required for this protocol could limit the noise sources. Additionally, we show how restricting the model to the single-excitation XY model allows the coherent phonon generation of even low $\alpha$ to be mitigated against. 

\section{\label{sec:expt_design}Experimental design}
\begin{figure}
    \centering
    \includegraphics[scale=0.82]{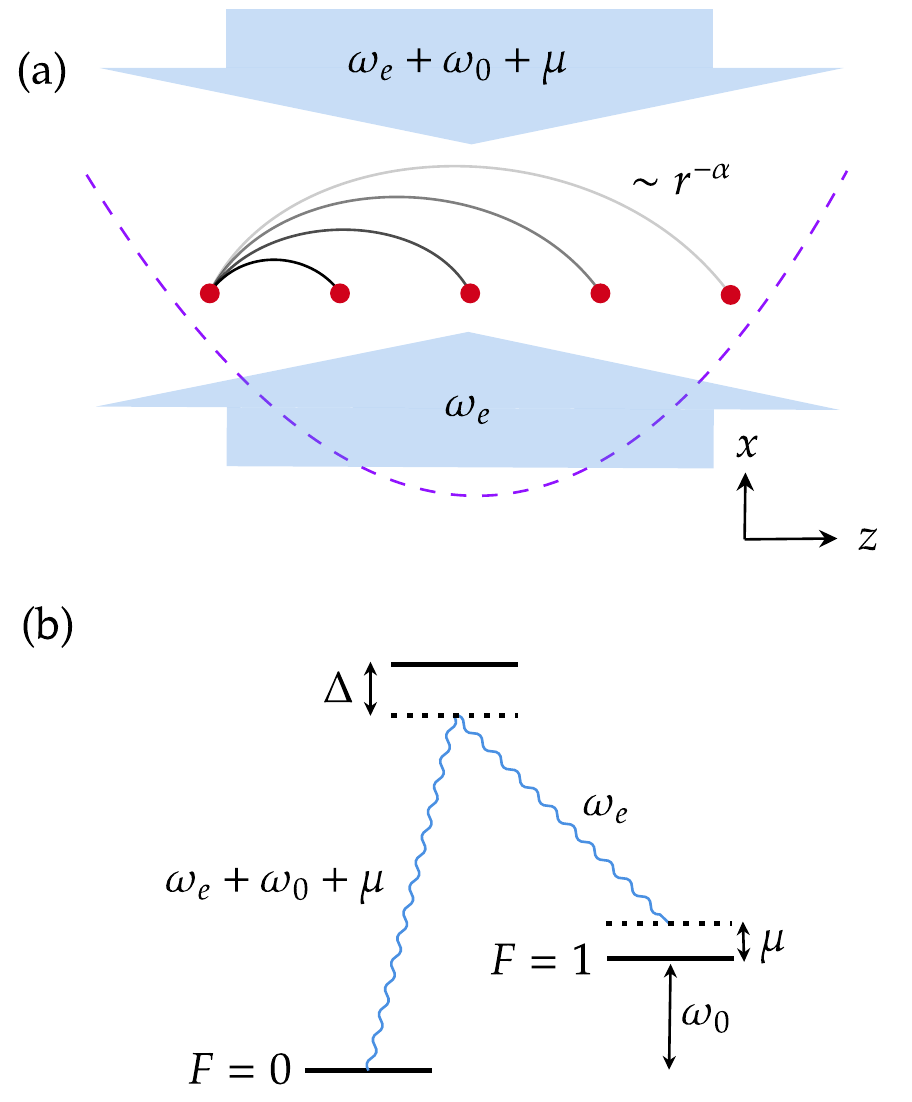}
    \caption{(a) The two-photon Raman transition is shown in blue with detuning $\Delta$ from the intermediate state and detuning $\mu$ from the excited state. (b) Illustration of chain of five $^{171}\textrm{Yb}^+$ ions (red dots) in the potential of an effective harmonic trap (purple dashed line) as described in the main text. The effective spin-spin interactions from just the first ion are depicted with the idealised $r^{-\alpha}$ power-law decay.}
    \label{fig:ion_chain_energy_levels}
\end{figure}

There are several ways to implement effective XY models in ion-trap chains~\cite{kiely_relationship_2017, monroe_programmable_2021}. The spin-dependent force between ions can be induced with a Mølmer-Sørensen scheme with only one sideband~\cite{wall_boson-mediated_2017}. A Raman transition is stimulated with bichromatic noncopropagating laser beams at the blue motional sidebands, so a frequency $\omega_0 + \mu$, where $\omega_0$ is the ion frequency, $\mu \approx \omega_c$, with $\omega_c$ being the transverse centre of mass phonon mode. The two off-resonant laser beams, with Rabi frequencies $g_1$ and $g_2$ are detuned by $\Delta$ from the excited intermediate level and the spin state transition is detuned by $\mu$ for all ions in the chain. The Rabi frequency is therefore $\Omega = \frac{g_1 g_2}{2\Delta}$ for every ion. The experimental platform we are considering is for $^{171}\textrm{Yb}^+$ ions, and the qubit states are encoded in the $F= \{0, 1\}$ hyperfine `clock' states of $S_{1/2}$, see Fig.~\ref{fig:ion_chain_energy_levels}. The interaction Hamiltonian for the system is 
\begin{multline}
    \label{eq:H_Interaction}
    H_I(t) = -\sum_{i,m} \frac{\Omega\eta_{i m}}{2} \Big(e^{-i (\omega_{\textrm{eff}}+\omega_m) t} a_m \sigma_i^- \\ 
    + e^{i(\omega_{\textrm{eff}} -\omega_m) t}a_m \sigma_i^+ + h.c.\Big),
\end{multline}
where $\omega_{\textrm{eff}} = \sqrt{\Omega^2 + \mu^2}$, $\eta_{i m} = b_{i m} \delta k \sqrt{\hbar / 2 M \omega_m}$ is the Lamb-Dicke parameter, $b_{i m}$ is the phonon mode transformation matrix, $\delta k$ is the wave vector difference of the counter-propagating Raman lasers, $\omega_m$ are the phonon mode frequencies, and $M$ is the mass of a single ion. Appendix~\ref{sec:derivation_interaction_hamiltonian} gives a detailed derivation of the interaction Hamiltonian. 
Appendix~\ref{sec:dyson_series} further derives the effective Hamiltonian by considering the first and second order terms of the Dyson series,
\begin{equation}
    \label{eq:xy_spin_hamiltonian}
    H_{XY} = \sum_{i\ne j} J_{ij}  \left( \sigma_j^x \sigma_i^x + \sigma_j^y \sigma_i^y \right) + \sum_{j} h_j \sigma_j^z.
\end{equation}
The coupling and single-site terms are 
\begin{align}
    \label{eq:coupling_calculation}
    J_{i j} &= \sum_{m} \frac{\Omega^2 \eta_{i m}\eta_{j m}\omega_m}{8(\omega_\textrm{eff}^2 - \omega_m^2)}, \\
    h_j &= \sum_{m} \frac{\Omega^2 \eta_{j m}^2 \omega_\textrm{eff} }{4(\omega_\textrm{eff}^2 - \omega_m^2)} \left( 2n + 1 \right),
    \label{eq:single_site_calculation}
\end{align}
where $n$ approximates the initial phonon number. 

The positions and the geometry of the ion chain determine the phonon modes $b_{i m}$ and the phonon mode frequencies $\omega_m$. At low temperatures, we can assume the ions are approximately in the configuration that minimises the potential energy of the ion chain. The potential $V$ contains a harmonic term from the trapping frequencies and a Coulomb repulsion term
\begin{multline}
    V = \frac{1}{2} \sum_{j=1}^{N} M (\omega_x^2 x_j^2 + \omega_y^2 y_j^2 + \omega_z^2 z_j^2) \\ + \frac{1}{2} \sum_{j=1}^N \sum_{i \ne j} \frac{e^2}{4\pi \epsilon_0 |\bm{r}_i - \bm{r}_j |},
\end{multline}
where $\bm{\omega} = (\omega_x, \omega_y, \omega_z)$ is the trapping frequency along the three spatial dimensions, and $\bm{r}_j = (x_j, y_j, z_j)$ is the position of particle $j$. The phonon modes, $b_{i m}$, and phonon mode frequencies, $\omega_m$, can be calculated by assuming small vibrations around the equilibrium positions of a linear chain and solving the equations of motion~\cite{fishman_structural_2008}. Once the phonon modes and frequencies are computed, the couplings $J_{i j}$ can be calculated for a specific detuning $\mu$ applied to all ions using Eq.~\eqref{eq:coupling_calculation}. 
In general, we have $\omega_x \approx \omega_y$ and $\omega_z < \omega_x,\omega_y$, such that a linear ion chain forms along the $z$ axis. Importantly, if $\omega_z$ is larger than a critical value for a given number of ions, $N$, relative to $\omega_x$ and $\omega_y$, there is a transition from a linear ion chain to a zig-zag chain~\cite{fishman_structural_2008, shimshoni_quantum_2011}. On the other hand, the $\omega_z$ trap frequency cannot be too small because it reduces axial confinement and the ions become well separated. This leads to phonon mode crowding, decreasing the accuracy of the approximation that the interactions can be described by a power-law decay as $\sim 1/r^\alpha$, and introduces a larger exponential decay factor $\beta$ in $J_{ij} \sim e^{-\beta r}/r^{\alpha}$. The $\omega_z$ trap frequency is determined by minimising $\beta$ while maintaining a linear ion chain. Some other experimental parameters do not change irrespective of $N$. For our simulations, these are: ion mass $M = 171$~\si{amu}, $\omega_x =  6\times 2\pi$~\si{MHz}, $\omega_y =  5\times 2\pi$~\si{MHz}, $\delta k = 4461.1 \times 10^{-9}$~\si{m^{-1}}, and $\Omega_\textrm{total} = 1 \times 2\pi $~\si{MHz}, where $\Omega_\textrm{total} = N \Omega$.

We define the $\alpha$ as the fit for $J_{1 j} \sim 1/r^\alpha$, the coupling strengths from an ion at the end of the chain to all the other ions along the chain. We motivate this choice in Appendix~\ref{sec:alpha_definition}. Changing the detuning $\mu$ alters the coupling strengths and therefore determines $\alpha$. 

The detuning cannot be arbitrarily low. The spin-spin interactions are mediated by spin-phonon interactions. The evolution is only described well by the effective spin-spin Hamiltonian if the phonons are only virtually excited. The maximum value of the first order term of the Dyson expansion, $\tilde{U}_1(t)$, derived in Appendix~\ref{sec:dyson_series}, occurs at $t_\textrm{max} = (1+2k) \pi /\Delta_m$, where $k \geq 0$ is an integer, $\Delta_m = \omega_\textrm{eff} - \omega_m$, and, after having applied the rotating wave approximation, we find   
\begin{equation}
    \tilde{U}_1 (t_\textrm{max}) = - \sum_{i,m} \frac{\Omega \eta_{im}}{\Delta_m}\left(a_m \sigma^+_i - a_m^\dagger \sigma_i^- \right).
\end{equation}
The coherent phonon generation for a single phonon mode is therefore negligible if
\begin{equation}
    \label{eq:detuning_condition}
    \Delta_m \gg \Omega \eta_{im},
\end{equation}
which is thus the condition for the pure spin-spin interaction model of Eq.~\eqref{eq:xy_spin_hamiltonian} to be valid. In this case the phonons are generally only virtual, and a very low proportion of real phonons are generated, we characterise this approximation in the following section. This is called the dispersive regime. 

In the resonant regime, where 
$\Delta_m \rightarrow 0$,
the first order terms $U_1(t)$ become important. In particular we have the secular term
\begin{equation}
    \lim_{\Delta_m \rightarrow 0}\tilde{U}_1 (t) = i \sum_{i,m} \frac{\Omega \eta_{im}}{2}\left(a_m \sigma^+_i + a_m^\dagger \sigma_i^- \right) t.
\end{equation}
The effective Hamiltonian is therefore the Jaynes-Cummings model, real phonons are generated, and we do not find the effective spin-spin model of Eq.~\eqref{eq:xy_spin_hamiltonian}.

\section{\label{sec:coherent_phonons}Coherent phonon generation in the XY model}
In this section we develop the theory to estimate the phonon production with a single phonon mode in the single excitation subspace. The ideas are applied to two phonon modes with similar results. We then consider the effect of the phonons on the effective XY Hamiltonian. Finally, higher excitation subspaces are considered to demonstrate that the effect on the XY model from coherent phonon generation is not limited to the single-excitation subspace.

\subsection{\label{sec:est_coherent_phonon_generation}Estimating coherent phonon generation in the single-excitation subspace}
\begin{figure*}[ht]
    \centering
    \includegraphics[scale=0.40]{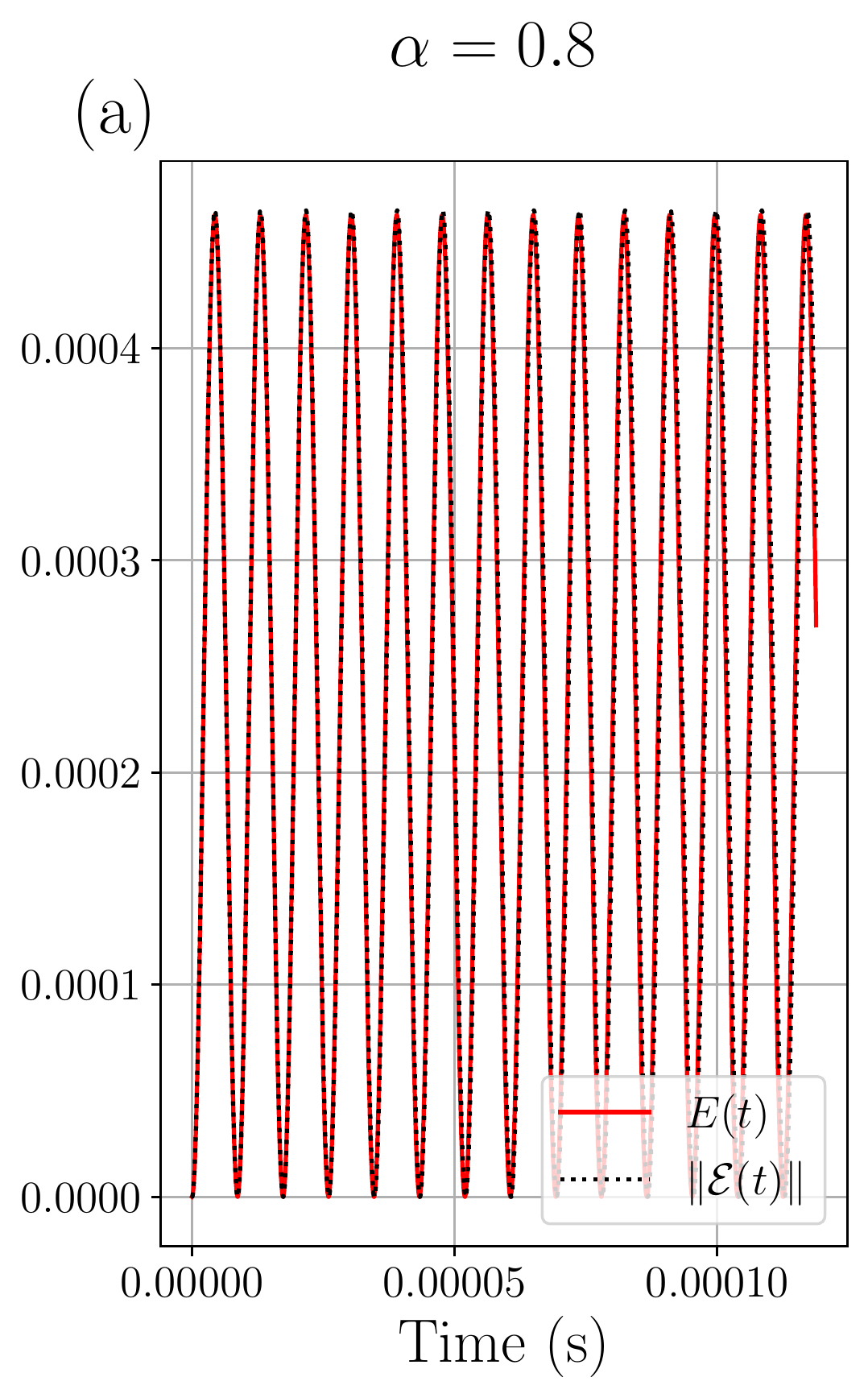}
    \includegraphics[scale=0.40]{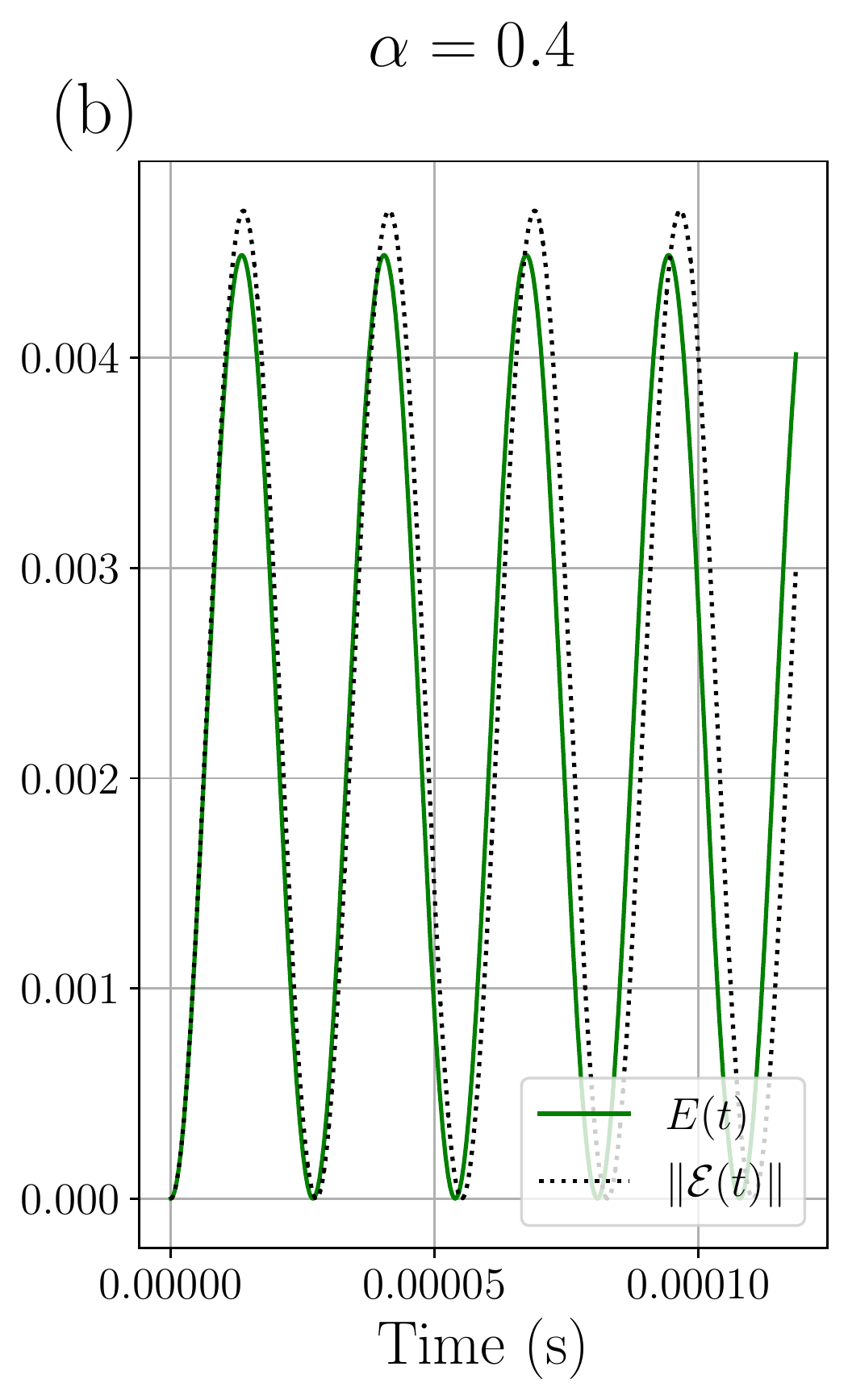}
    \includegraphics[scale=0.40]{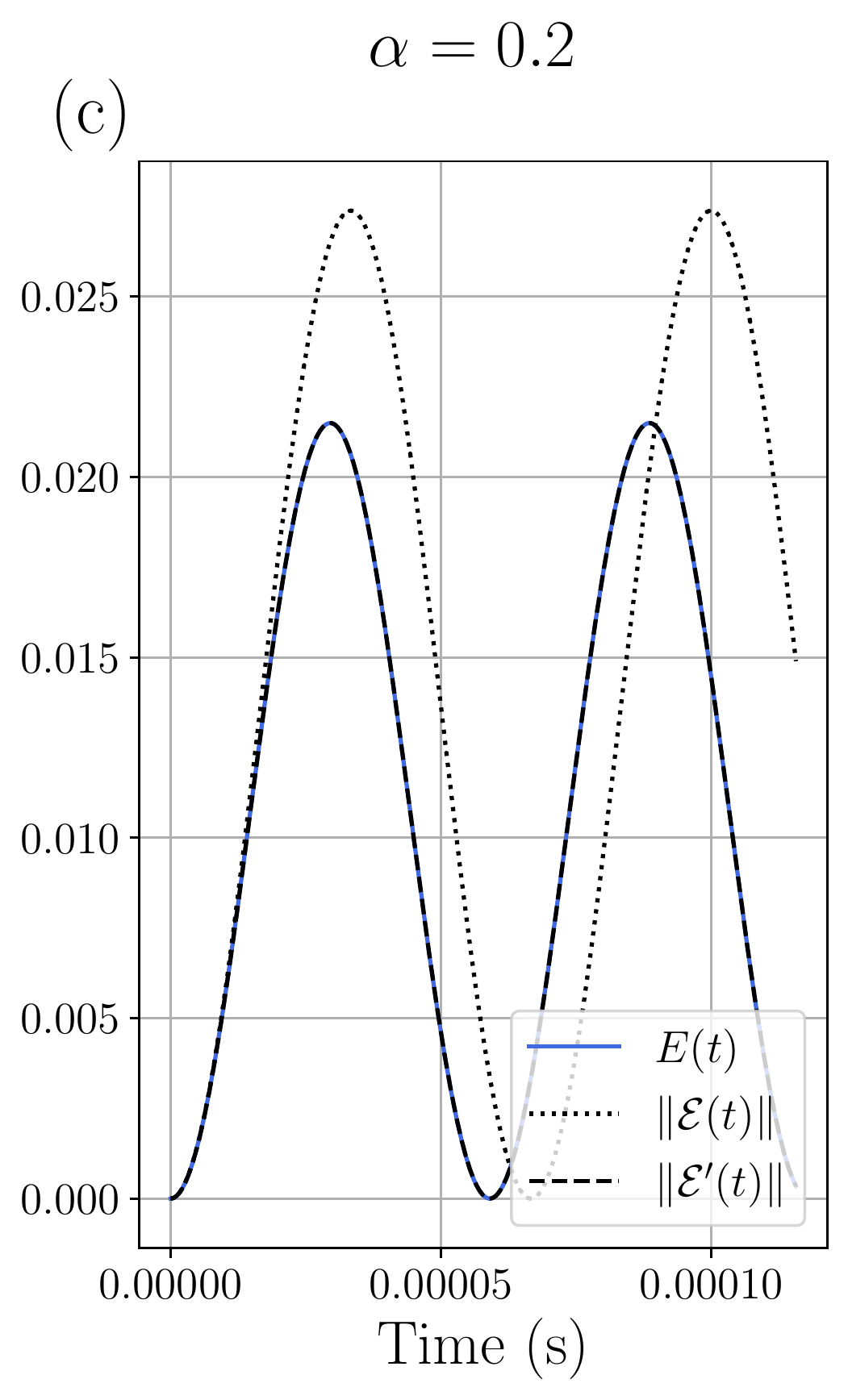}
    \includegraphics[scale=0.40]{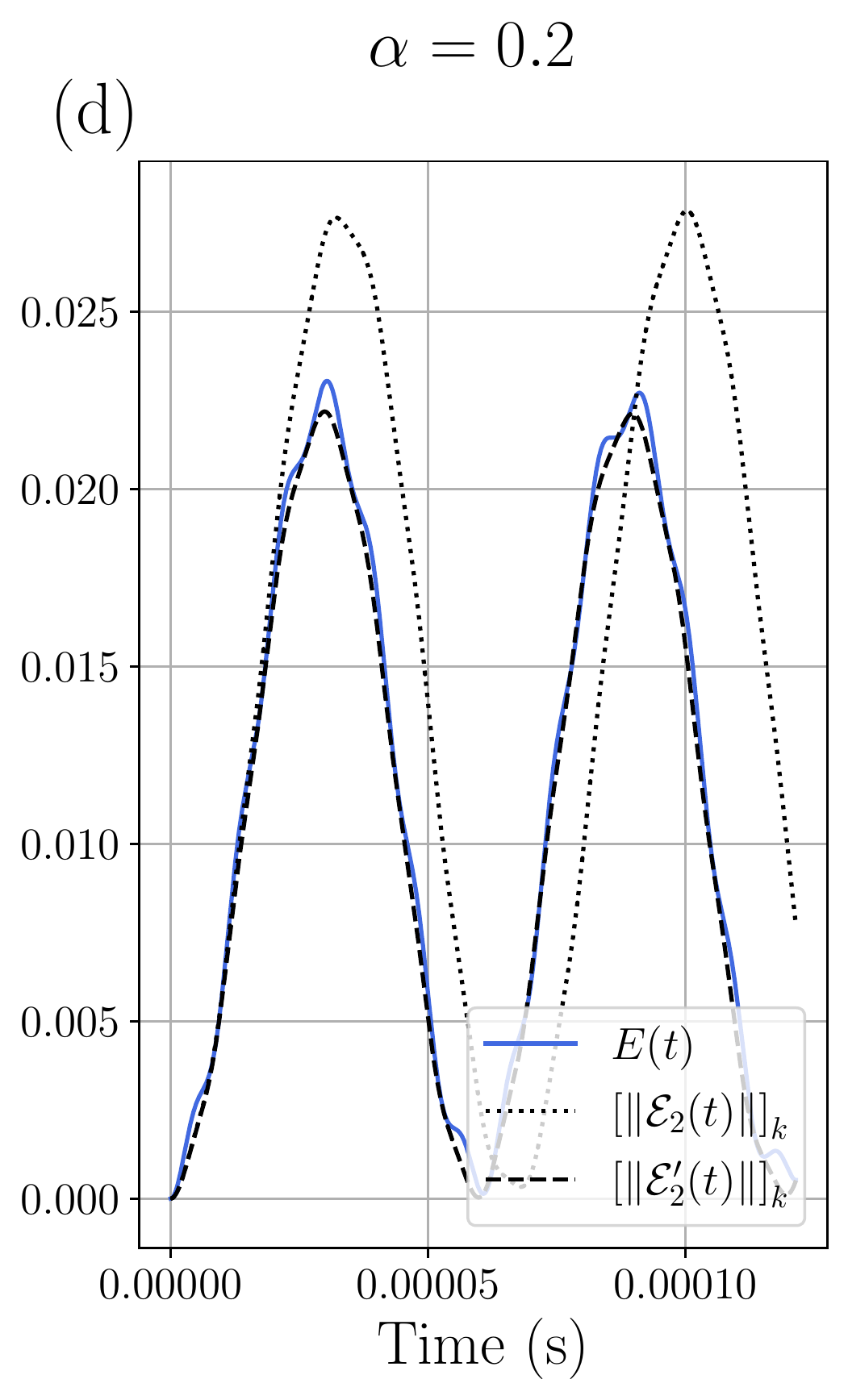}
    \caption{(a)-(c) For a single phonon mode, the transverse centre of mass mode $\omega_c = 36.774848~\textrm{MHz}$, the leakage operator $\mathcal{E}(t)$ is compared to the overlap of simulated full system dynamics $H_I(t)$ from Eq.~\eqref{eq:H_Interaction} with all spins in ground state, $E(t)$, with $N=10$ ions for (a) $\alpha=0.8$, (b) $\alpha=0.4$, and (c) $\alpha=0.2$. In (c), $\Vert \mathcal{E}^\prime (t) \Vert$ is additionally plotted, which is with $\omega_\textrm{eff}^\prime = 1.000321 \omega_\textrm{eff}$ giving a perfect fit for $E(t)$. (d) The average leakage from the most significant two phonon modes, $\left[ \Vert \mathcal{E}_2(t) \Vert\right]_{k}$ as defined in Eq.~\eqref{eq:two_phonon_mode_leakage}, for $\alpha=0.2$ and $N=10$ ions is compared to the error due to coherent phonon production $E(t)$. $\left[ \Vert \mathcal{E}_2^\prime(t) \Vert\right]_{k}$ is also shown, which is the leakage for the most significant two phonon modes with $\omega_\textrm{eff}^\prime = 1.000306 \omega_\textrm{eff}$.}
    \label{fig:leakage_plots}
\end{figure*}

Using the Dyson series, the evolution of the system is 
\begin{multline}
    \rho (t) = \left[\mathds{1} + \tilde{U}_1(t) + \tilde{U}_2(t) + \dots \right]\rho(0) \\ \left[\mathds{1} + \tilde{U}_1^\dagger (t) + \tilde{U}_2^\dagger (t) + \dots\right],
\end{multline}
where $\tilde{U}_n$ indicates the $n$th order Dyson term with applied rotating wave approximations, as in the derivation of the effective Hamiltonian $H_{XY}$, see Appendix~\ref{sec:dyson_series}. The evolution of the spin subsystem is $\rho_\textrm{sp} (t) =  \mathrm{Tr}_\textrm{ph}\left[\rho (t) \right]$. $\tilde{U}_1(t)$ contains a term with $a$ and a term with $a^\dagger$, changing the phonon number by exactly one in both terms. $\tilde{U}_2(t)$ only acts on the spin subsystem. The partial trace over the phonons therefore gives 0 for the following terms: $ \mathds{1} \rho(0)\tilde{U}_1^\dagger(t)$, $\tilde{U}_1(t) \rho(0)\mathds{1}$, $\tilde{U}_2(t) \rho(0)\tilde{U}_1^\dagger (t)$, and $\tilde{U}_1(t) \rho(0)\tilde{U}_2^\dagger(t)$.
Thus, the spin subsystem evolution is 
\begin{align}
     \rho_\textrm{sp} (t) &\approx \mathcal{N}(t) e^{-i H_{XY} t} \rho_\textrm{sp}(0)  e^{i H_{XY} t}  + \mathcal{E}(t),
     \label{eq:spin_subsystem_evolution}
\end{align}
where the approximation is valid in the same regime as for the effective Hamiltonian $H_{XY}$ in Eq.~\eqref{eq:xy_spin_hamiltonian}, i.e. $\mathds{1}\otimes e^{-i H_{XY} t} \approx \mathds{1} + \tilde{U}_2(t)$, and we have defined a leakage operator,
\begin{equation}
    \mathcal{E}(t) = \mathrm{Tr}_\textrm{ph}\left[  \tilde{U}_1(t) \rho(0)\tilde{U}_1^\dagger (t) \right],
\end{equation}
that quantifies the error from the XY spin-spin model due to first-order coherent phonon generation. In order to preserve the norm of the partial trace, we have also defined $\mathcal{N}(t) = 1- \textrm{Tr}\left[\mathcal{E}(t)\right]$. The initial state is $\rho(0) = |\psi_0 \rangle\langle \psi_0|$, with $|\psi_0\rangle = |0\rangle_{\textrm{ph}}|10\dots0\rangle_{\textrm{sp}}$, and 
\begin{multline}
    \label{eq:U_1_RWA_1_phonon_mode}
    \tilde{U}_1(t) = i \sum_{m,j} \frac{\Omega \eta_{j,m}}{2} \big(\alpha_{m}(0,1;t) a_m \sigma_j^+ \\ + \alpha_{m}(1,0;t) a_m^\dagger \sigma_j^- \big),
\end{multline}
where the sum is over the phonon modes $m$ and ions $j$, $\alpha_m(1,0;t)$ is a prefactor dependent on the phonon mode and time, as derived in Appendix~\ref{sec:dyson_series}. We therefore find
\begin{align}
    \mathcal{E}(t) &= \frac{1}{4} \sum_{m,l} \Omega^2 \eta_{1,m} \eta_{1,l} \alpha_{m}(1,0;t) \alpha^*_{l}(1,0;t) |\bm{0}\rangle \langle\bm{0}|,
\end{align}
where $|\bm{0}\rangle$ is the state with all spins in the ground state. This state space is outside the XY spin-spin model, which, assuming a perfect model and given the initial state, should remain entirely within the single-excitation subspace. This further qualifies $\mathcal{E}(t)$ as a leakage from the XY spin-spin model. The state of the spins after time $t$ can therefore be approximated by a linear combination of inside the desired excitation subspace, 
$\rho_{XY}(t)$, and outside, $\rho_{\mathcal{E}}(t)$,
\begin{equation}
    \label{eq:state_leakage_proportion}
    \rho(t) \approx \mathcal{N}(t) \rho_{XY}(t) + \Vert \mathcal{E}(t) \Vert \rho_{\mathcal{E}}(t) ,
\end{equation}
where $\rho_{\mathcal{E}}(t) = |\bm{0}\rangle\langle\bm{0}|$ for the single-excitation subspace, and the approximation is due to defining $\rho_{XY}(t)$ as the exact evolution of the state in the XY model rather than due to the interaction Hamiltonian of Eq.~\eqref{eq:H_Interaction}. Considering only the largest phonon contribution, the transverse centre of mass mode labelled $c$, and using $\alpha_{c}(1,0;t)$ as defined in Appendix~\ref{sec:dyson_series}, gives
\begin{equation} 
    \|\mathcal{E}(t)\| = \frac{\Omega^2 \eta_{1,c}^2\left(1-\cos(\Delta_c t)\right)}{2\Delta_c^2}.
\end{equation}
The phonon production leads to a maximum error of $\Omega^2 \eta_{1,c}^2/\Delta_c^2$ at times $t_\textrm{max} = (1+2 k)\pi/\Delta_c$, for integer $k\ge 0$, which again gives the condition of Eq.~\eqref{eq:detuning_condition}.
With this approximation, the fidelity of the general Hamiltonian with the XY model is
\begin{align}
    F(t) &=  \textrm{Tr}\left[\rho(t) \mathds{1}\otimes\rho_{XY}(t)\right] 
    \label{eq:fidelity_definition}\\
    &\approx 1-\|\mathcal{E}(t)\|. \label{eq:fidelity_xy_leakage}
\end{align}
We can investigate how well the leakage operator captures the error due to coherent phonon production by defining 
\begin{equation}
    E(t) = |\langle \bm{0}|\mathrm{Tr}_{\textrm{ph}}\left[U(t) \rho(0) U^\dagger(t)\right] |\bm{0}\rangle|,
\end{equation}
where the state at time $t$ is due to the full evolution of the interaction Hamiltonian of Eq.~\eqref{eq:H_Interaction}, given by $U(t)$, of the initial state $\rho(0)$. 
In Fig.~\ref{fig:leakage_plots}, we find that $E(t) \approx \Vert \mathcal{E}(t) \Vert$, and that this approximation becomes more accurate as $\alpha$ increases, when the detuning $\mu$ increases and the interaction becomes less long range. However, even as the approximation becomes less accurate, $\Vert \mathcal{E}(t) \Vert$ overestimates the leakage, and can therefore be considered a bound on coherent phonon generation. 

The coherent phonon generation can also be quantified by the phonon mode occupation number. For high $\alpha$, only virtual phonons are excited. However, for lower $\alpha$, such as $\alpha=0.2$, we find that the phonon occupation number perfectly aligns with our leakage $E(t)$, and only a single phonon is generated from our initial state, see Fig.~\ref{fig:phonon_generation_higher_subspaces}(a). The leakage operator may therefore also be defined as the single phonon generation operator – as we show in Section~\ref{sec:higher_subspaces}, this also applies to higher excitation subspaces.

This analysis can be extended to two phonon modes with similar results. In particular, the leakage is still dominated by the transverse centre of mass mode for $\alpha \ge 0.2$, however, the additional mode contributes. For the transverse centre of mass mode and the next most significant phonon mode, we find the leakage
\begin{multline}
    \label{eq:two_phonon_mode_leakage}
    \left[ \Vert \mathcal{E}_2(t) \Vert\right]_{k} = \frac{\Omega^2}{2 N} \sum_{k=1}^{N}\Bigg[\frac{\eta_{k,1}^2}{\Delta_1^2}\left(1-\cos(\Delta_1 t)\right)\\ 
    + \frac{\eta_{k,1}\eta_{k,2}}{\Delta_1\Delta_2}\left(1-\cos(\Delta_1 t)-\cos(\Delta_2 t) + \cos((\omega_1-\omega_2) t)\right) 
    \\ + \frac{\eta_{k,2}^2}{\Delta_2^2}\left(1-\cos(\Delta_2 t)\right)\Bigg],
\end{multline}
where the phonon modes are labelled $1$ and $2$. The average over the ion $k$ has been computed because the position in the chain determines $\eta_{k,2}$ for ion $k$ interacting with phonon mode $2$. The second phonon mode gives the same conclusions as for a single phonon mode, see Fig.~\ref{fig:leakage_plots}(d).

\subsection{\label{sec:effect_coherent_phonons}Effect of coherent phonons on the single-excitation XY model}
The question is whether this coherent phonon generation affects the effective XY model of Eq.~\eqref{eq:xy_spin_hamiltonian} beyond the requirement of stroboscopic measurement for maximum fidelity. Naively assuming the model remains as derived, we find a significant dephasing effect for low $\alpha$ over the coherence time of a typical ion motional state ($\sim 10~\textrm{ms}$~\cite{monroe_programmable_2021}).
However, careful treatment of the system shows the XY model remains, only with slightly decreased interaction strength. 

In order to show that the simplified XY Hamiltonian of Eq.~\eqref{eq:xy_spin_hamiltonian} is an accurate model, we compute the full dynamics of up to four phonons with a single phonon mode and $N=10$ ions. Simulations of the general interaction Hamiltonian of Eq.~\eqref{eq:H_Interaction}, but with only the centre-of-mass transverse phonon mode, are compared with the XY model for the ions of Eq.~\eqref{eq:xy_spin_hamiltonian}.
Fidelity, $F(t)$, is the figure of merit for the accuracy of the XY model. The fidelity of the XY model is defined in Eq.~\eqref{eq:fidelity_definition}, but where the full evolution has only a single phonon mode. The initial state for both the full evolution and the XY model is 0 phonons and only the first ion excited.

As noted previously, for the initial state considered and $\alpha \geq 0.2$, we find only a single phonon is excited, see Fig.~\ref{fig:phonon_generation_higher_subspaces}(a). It is therefore sufficient to consider the leakage operator as capturing the phonon production. However, $\Vert \mathcal{E}(t) \Vert$ overestimates the amplitude of $E(t)$ with a lower frequency. We can correct this model by changing the $\Delta_c$ in $\Vert \mathcal{E}(t) \Vert$. 

We can explain this through additional higher-order secular terms in the Dyson series becoming more relevant as the coherent phonon generation becomes stronger. Physically, the increased amplitude of real phonon generation is captured by the magnitude of the non-secular first order Dyson series terms. While the system contains phonons, albeit with small amplitude, a fraction of the spin state lies outside the target XY model subspace. For example, for the single-excitation subspace the spin state becomes $|\bm{0}\rangle \langle \bm{0}|$, which does not evolve, and the evolution of the XY model is effectively slowed down. This effect requires the spins going through the state of $|\bm{0}\rangle \langle \bm{0}|$, i.e. a process that has two phonons (and therefore two virtual spins as well). This term would only be included in the fourth order Dyson series term and is thus not captured by our effective Hamiltonian.

We account for this effect at low detuning by considering an effective shift in the ion frequency $\omega_0$. The shift in ionic frequency, $\omega^\prime_0 = \omega_0 + \delta \omega$, leads to a shift in the effective frequency, $\omega_\textrm{eff}^\prime = r \omega_\textrm{eff}$, further leading to an updated leakage, $\Vert \mathcal{E}^\prime(t) \Vert$ – shown in Fig.~\ref{fig:leakage_plots}(c) for $N=10$ ions.

\begin{figure*}[ht]
    \centering
    \includegraphics[scale=0.4]{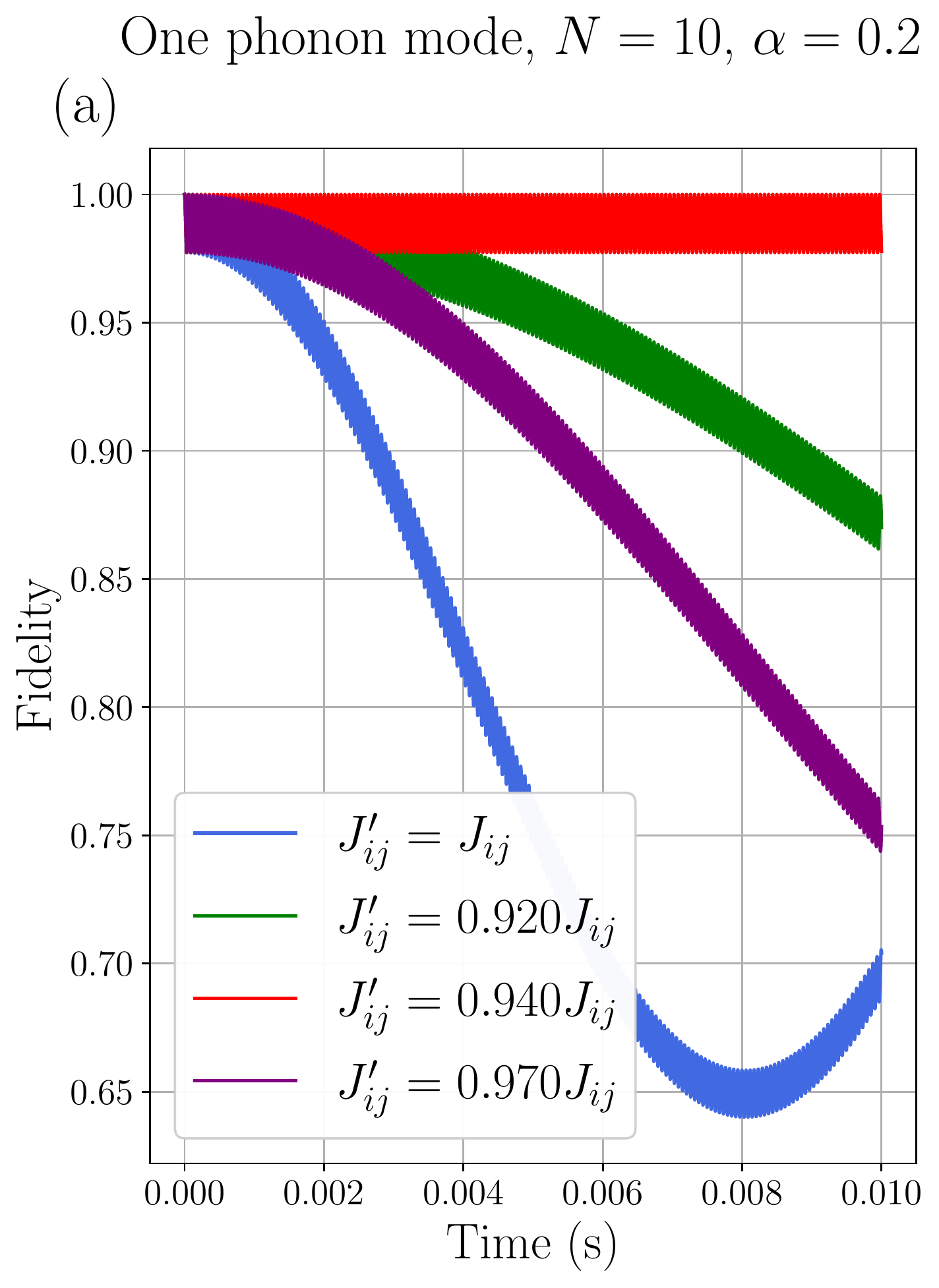}
    \includegraphics[scale=0.4]{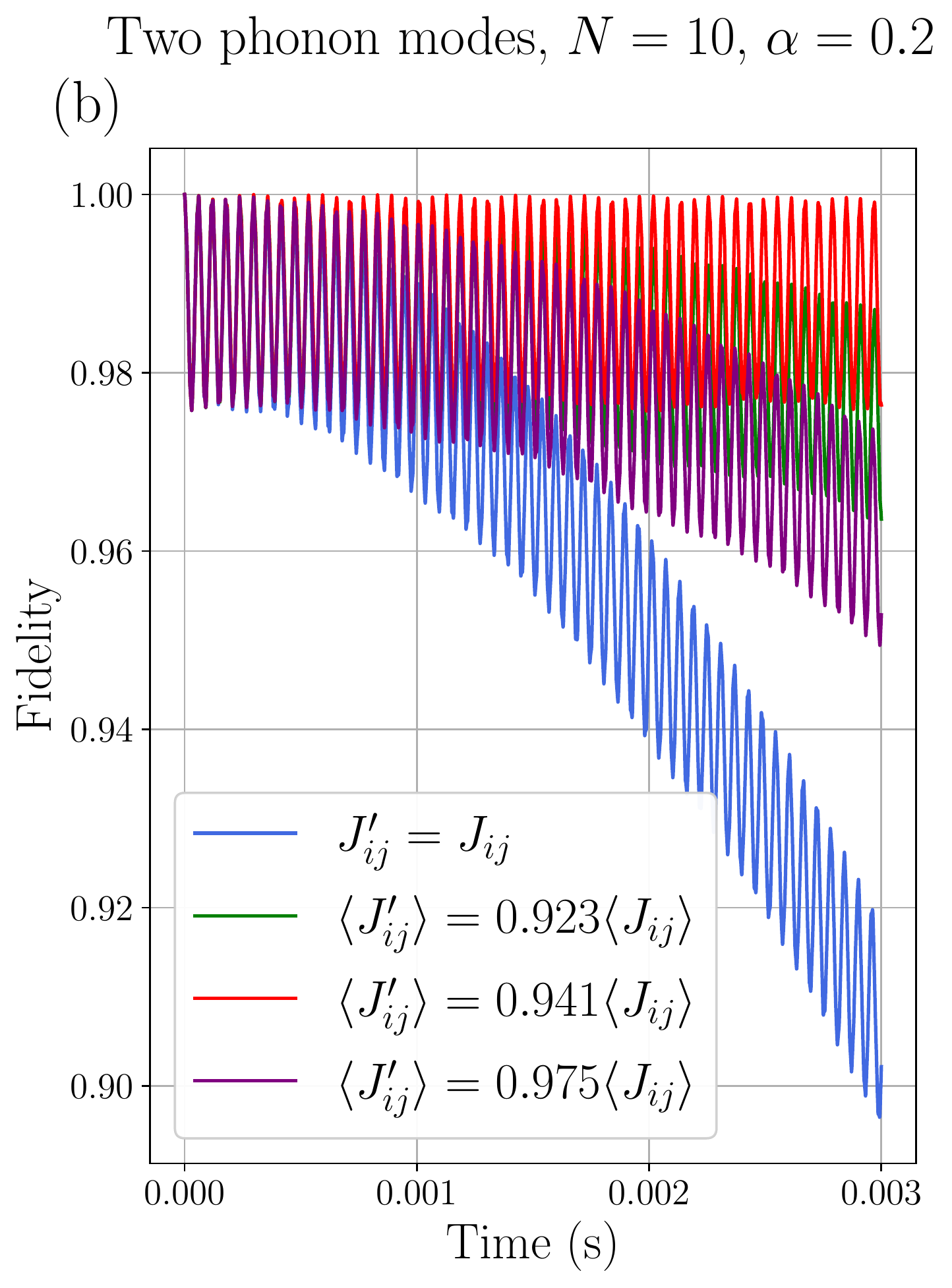}
    \includegraphics[scale=0.4]{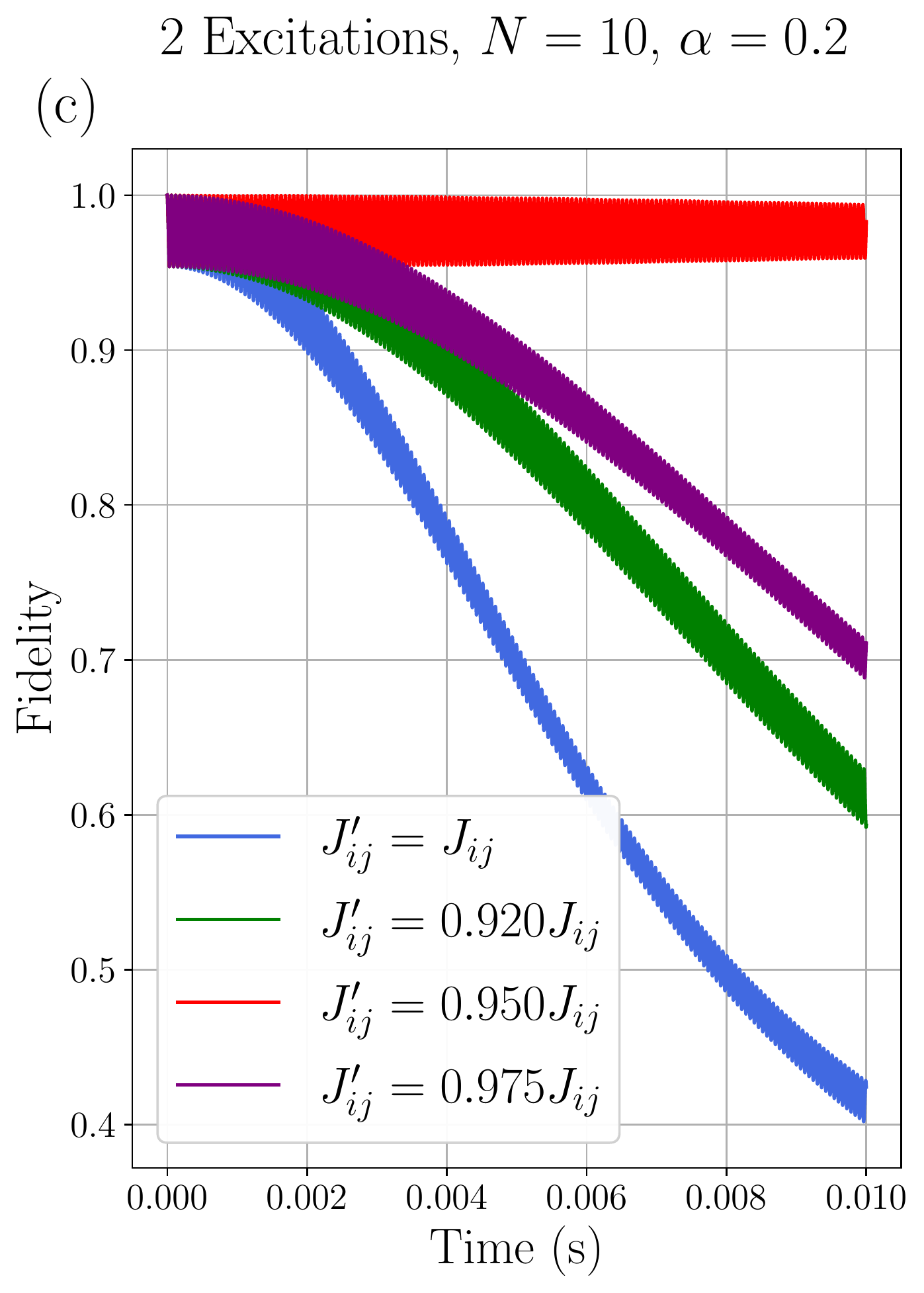}
    \caption{(a)-(b) For the initial state of no phonons and a single excited ion, the fidelity of the ion subspace is computed for the evolution due to the full interaction Hamiltonian of Eq.~\eqref{eq:H_Interaction} for: (a) a single phonon mode with the XY model of Eq.~\eqref{eq:xy_spin_hamiltonian} for various $r$; (b) two phonon modes, where $\langle \cdot \rangle$ indicates the average of the interactions strengths – the precise scaling is dependent on the specific interaction $J_{ij}^\prime$. The fidelity oscillates fast and regularly at frequency $\Delta_c^\prime$ with peaks and troughs separated by $\Omega^2\eta^2/\Delta_c^{\prime 2}$, as derived in Section~\ref{sec:est_coherent_phonon_generation}. The various $r$ give scaled coupling strengths, for (a) $0.957 J_{ij}$ (red), $0.975 J_{ij}$ (green) and $r=1$ simply gives $J_{ij}$ (blue); and (b) $0.960 J_{ij}$ (red), $0.975 J_{ij}$ (green), and $J_{ij}$ (blue). (c) shows the fidelity of the XY model for an initial state with two ion excitations and no phonons and a single phonon mode is simulated with $0.967 J_{ij}$ (red), $0.957 J_{ij}$ (green), and $J_{ij}$ (blue).}
    \label{fig:fidelity_xy_models}
\end{figure*}

Increasing $\omega_\textrm{eff}$ decreases the interaction strengths $J_{ij}$ of the effective XY model. However, in the case that the unitary evolution from the coherent phonons, $\tilde{U}_1(t)$, vanishes, which is also when $E(t) = 0$, there must be a pure XY model with the original $\omega_\textrm{eff}$. Hence, we cannot simply use $\omega_\textrm{eff}^\prime$ to compute $J^\prime_{ij}$. Since the regime that $\Delta$ is large compared to $\Omega\eta$ still applies, we can instead use a time averaged ion frequency $(1+r)\omega_\textrm{eff}/2$ for the interaction strength as the effective frequency oscillates between $\omega_\textrm{eff}$ and $\omega_\textrm{eff}^\prime$. Thus, we find $J^\prime_{ij} = 0.940 J_{ij}$ for $N=10$ ions, which is precisely what is found numerically by simulations, see Fig.~\ref{fig:fidelity_xy_models}(a). This order of variation in coupling strength should be largely due to fourth order contributions to the Dyson series, the coupling strength difference between the second order effective Hamiltonian and fourth order terms is $\sim (\eta \Omega / 2 \Delta )^2$, which is $\sim 10^{-2}$ for $\alpha =0.2$. There are also additional interaction paths at fourth order, i.e. the spin-spin interaction is now mediated by two virtual phonon-spin-phonon diagrams and any spin can be the mediator for the interaction path. 

Remarkably, even for low $\alpha$, we have shown that the XY model is preserved – only that the interaction strengths between the ions have been decreased. The coherent phonon production of one phonon mode does not dephase the XY model for the initial state that we consider in this protocol and the model error is therefore well characterised by the leakage defined in the previous section. 

Introducing a second phonon mode slightly changes the $\omega_\textrm{eff}^\prime$ that we find. The second phonon mode is significantly more detuned, thus there is essentially no coherent phonon generation for this mode and no frequency shift, $\omega_\textrm{eff}^\prime \approx \omega_\textrm{eff}$ for the second phonon mode. Therefore the total effect of the ratio $J^\prime / J$ is slightly reduced because the new virtual phonon path, the second phonon mode, does not have such a shifted effective frequency. We use the same method as with a single phonon mode to find $\omega_\textrm{eff}^\prime$ – using the coherent phonon generation from the leakage, as shown in Fig.~\ref{fig:leakage_plots}(d). For $N=10$ ions, we find that $\omega_\textrm{eff}^\prime = 1.000306 \omega_\textrm{eff}$ (compared to $\omega_\textrm{eff}^\prime = 1.000321 \omega_\textrm{eff}$ for a single phonon mode). Averaging over all interactions between ions, this leads to a shift of $J_{ij}^\prime = 0.941 J_{ij}$ (compared to $J_{ij}^\prime = 0.940 J_{ij}$ for a single phonon mode). This is confirmed with simulations of initial state evolutions, see Fig.~\ref{fig:fidelity_xy_models}(b), where each $J_{ij}^\prime$ is shifted by different factors (with average 0.941) to reproduce a perfect fidelity XY model at long times.

\subsection{\label{sec:higher_subspaces}Higher-excitation subspaces}
\begin{figure*}
    \centering
    \includegraphics[scale=0.4]{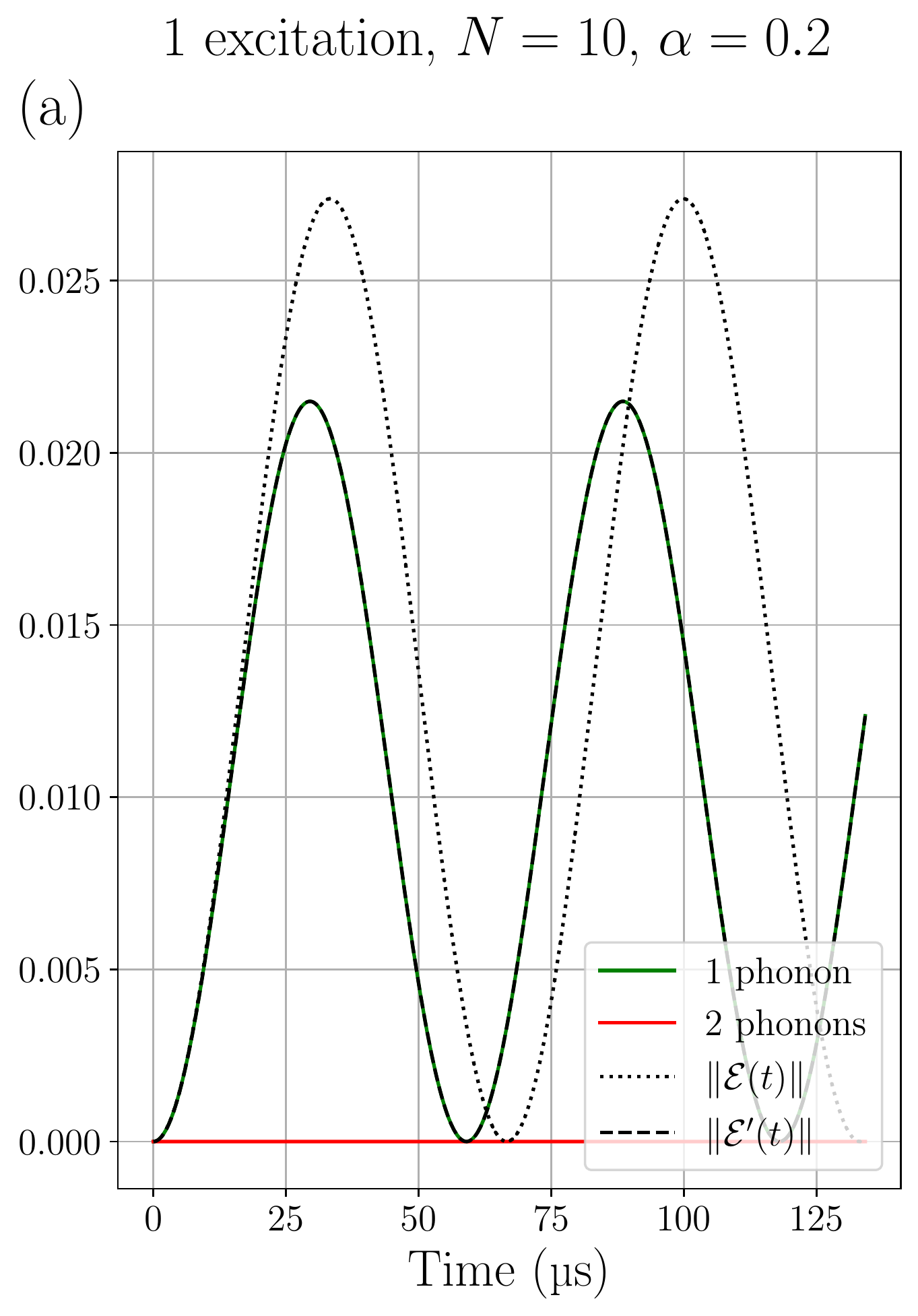}
    \includegraphics[scale=0.4]{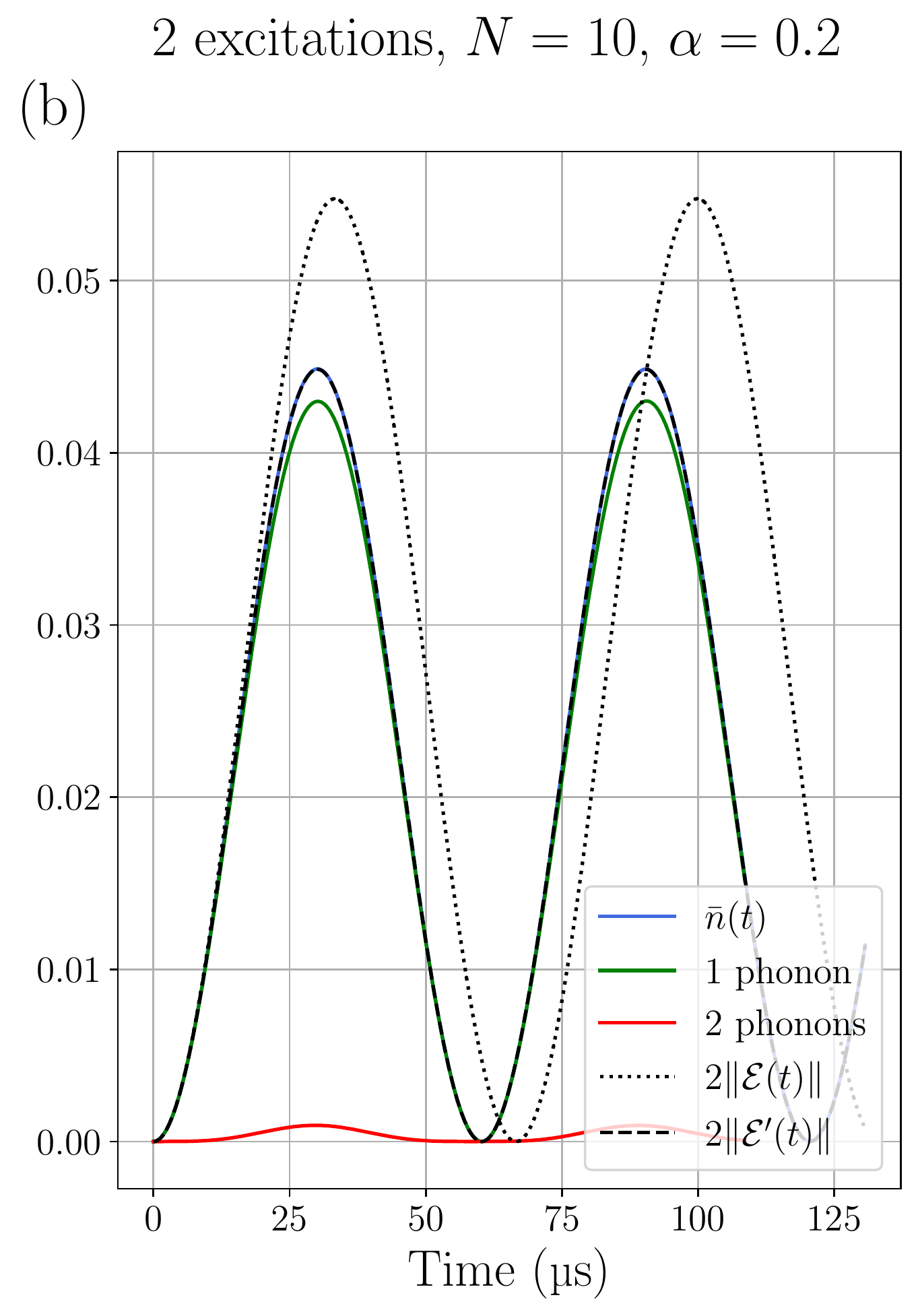}
    \includegraphics[scale=0.4]{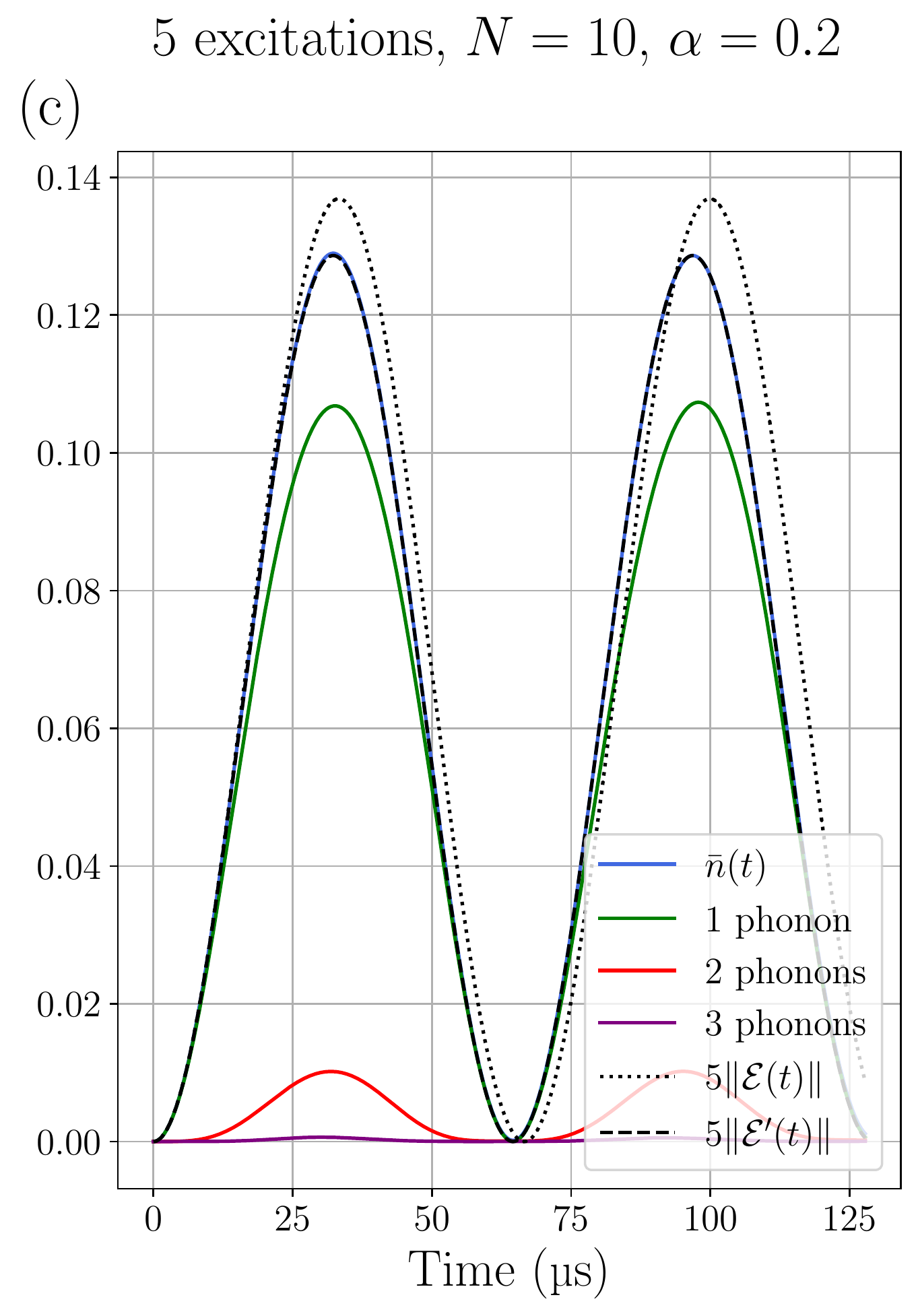}    
    \caption{Comparing the analytical $s \Vert \mathcal{E}(t) \Vert$ with the simulated phonon generation for a single phonon mode and $N=10$ ions, where $s$ is the initial number of spin excitations in the ion chain. The number of phonons is defined for $n$ phonons as $n \langle n | \hat{n} \textrm{Tr}_\textrm{sp} \left[ \rho(t)   \right] |n\rangle $ and $\rho(t)$ is from the evolution due to $H_I$, from Eq.~\eqref{eq:H_Interaction} with a single phonon mode. The plots show the phonon generation for a chain of various initial states: (a) shows a chain with one initial excitation, the simulated single phonon occupation number is well characterised by the leakage $\Vert \mathcal{E}^\prime(t) \Vert$ with $\omega_\textrm{eff}^\prime = 1.000321\omega_\textrm{eff}$; (b) shows a chain with two initial excitations, the simulated phonon number $\bar{n}(t)$ is well characterised by the leakage $2 \Vert \mathcal{E}^\prime (t)\Vert$ with $\omega_\textrm{eff}^\prime = 1.000262\omega_\textrm{eff}$; (c) shows a chain with five initial excitations, the simulated phonon number $\bar{n}(t)$ is well characterised by the leakage $5 \Vert \mathcal{E}^\prime(t) \Vert$ with $\omega_\textrm{eff}^\prime = 1.0000788\omega_\textrm{eff}$.}
    \label{fig:phonon_generation_higher_subspaces}
\end{figure*}

The previous analysis only strictly applies for an initial state with one excitation and no phonons, which is the case for optimal spatial search and the quantum state transfer protocol. To generalise the leakage to higher subspaces, we consider the phonon occupation number $\bar{n}(t) =   \textrm{Tr} \left[ \hat{n} \rho_\textrm{ph}(t)  \right]$, 
where $\rho_\textrm{ph}(t) = \textrm{Tr}_\textrm{sp} \left[ \rho(t) \right]$, numerically we truncate the occupation at the initial number of excitations. In the single-excitation subspace we found $\bar{n}(t) \approx \Vert \mathcal{E}(t) \Vert$. In higher-excitation subspaces, we show that the coherent phonon generation is well characterised by $\bar{n}(t) \approx s  \Vert \mathcal{E}(t) \Vert$, where $s$ is the initial number of excitations. This is the case despite the increased average phonon number for the increased number of initial excitations, see Fig.~\ref{fig:phonon_generation_higher_subspaces}. 

As with the single excitation subspace, $s  \Vert \mathcal{E}^\prime (t) \Vert$ can be made to match $\bar{n}(t)$ by slightly increasing the effective frequency $\omega_\textrm{eff}^\prime$. Hence, $J_{ij}^\prime$ is found using the same method as previously described. In the case of four initial excitations for $N=10$ ions, we find $J_{ij}^\prime = 0.988 J_{ij}$, and for two initial excitations, we find $J_{ij}^\prime = 0.967 J_{ij}$, see Fig.~\ref{fig:fidelity_xy_models}(c). The difference between $\omega_\textrm{eff}^\prime$ and $\omega_\textrm{eff}$ decreases as the number of initial excitations increase. This is due to the fourth order Dyson series term that it stems from: the spin-spin interactions are mediated by a two phonon process that now includes two virtual spins. The number of routes that contain spins that can be used decreases as the number of excitations in the chain increases.

\section{\label{sec:protocol}Transfer protocol}
The interaction model we are considering is the XY spin chain in the single-excitation subspace,
\begin{equation}
    \label{eq:walk}
    H = \sum_{i<j} J_{ij} \left(|i\rangle\langle j| + |j\rangle\langle i| \right),
\end{equation}
where the basis states, $|j\rangle$, are a single spin excitation being at the site $j$ -- we can consider all spins being down apart from a single spin at site $j$ being up. Here, we consider the one-dimensional open spin chain with long range interactions, dependent on the distance between the spins, $J_{ij} \propto  |j-i|^{-\alpha}$,
where the range of the interaction is parameterised by $\alpha$. 

The entire chain is initialised in the down state apart from the sender site of the spin chain. The sender site is in an arbitrary quantum state $|\psi\rangle = 
\alpha |0\rangle + \beta|1\rangle$. In the zero-excitation subspace, when the sender site is in state $|0\rangle$, as is every other spin, the evolution is trivial and we remain in this state -- an eigenstate of the Hamiltonian. We therefore only have to consider the case that the sender site is in the state $|1\rangle$, and we are in the single-excitation subspace.

The protocol has three simple steps. First, assuming we are transferring the quantum state between the ends of the chain, we initialise the chain into the state $|\psi\rangle \otimes |00 \dots \rangle$. We then evolve the chain under the Hamiltonian 
\begin{equation}
    H_\textrm{s} = \gamma H + |w\rangle\langle w| + |f\rangle \langle f|,
\end{equation}
where $\gamma$ is the relative strength of the interactions strengths between the sites to the strength of the marking field at specific sites, $w$ is the initial site (e.g. 1), and $f$ is the final site (e.g. $n$). Finally, we switch off the Hamiltonian after time $T = \pi \sqrt{\frac{n}{2}}$ such that there is a high fidelity that we have the state $|\psi\rangle$ at the site $f$. 

A similar protocol has been considered before in a very different setting~\cite{chakraborty_spatial_2016}, for the case of asymptotic Erdös-Renyi random graphs, and without considering a physically realisable model for implementation. Our protocol utilises the natural dynamics of systems that are already realisable, with experimental demonstrations in various platforms: dipolar crystals~\cite{micheli_toolbox_2006,yan_observation_2013}, Rydberg arrays~\cite{browaeys_many-body_2020}, and ion traps~\cite{kim_entanglement_2009, lanyon_efficient_2017,richerme_non-local_2014,friis_observation_2018,monroe_programmable_2021}, as we describe in Section~\ref{sec:expt_design}.

The protocol works in the regime where optimal spatial search is possible in the long-range interaction setting~\cite{lewis_optimal_2021}. Although the fidelity decreases as the interaction strength increases, these protocols will allow quantum state transfer for $\alpha<1.5$. 

In the following, we show that after time $T= \pi \sqrt{\frac{n}{2}}$ there is high fidelity of quantum state transfer from site $w$ to site $f$, which increases asymptotically as $n \rightarrow \infty$. This analysis assumes the Hamiltonian can be accurately described by a degenerate subspace of the three most significant states. The degeneracy of the subspace is a good approximation for well-chosen $\gamma$, however, it does therefore rely on the same spectral gap conditions as optimal spatial search~\cite{chakraborty_optimality_2020, lewis_optimal_2021}. 

For a particular choice of $\gamma$, in graphs where optimal spatial search is possible~\cite{chakraborty_optimality_2020}, there are three eigenenergies corresponding to $|w\rangle, |f\rangle$, and the state $|\phi_n 
\rangle$ that are separated from the rest of the spectrum, where $|\phi_n 
\rangle$ is the eigenstate of $H$ with the largest eigenvalue $\lambda_n$. For this particular $\gamma$, the Hamiltonian can be rewritten with only the significant terms,
\begin{equation}
    H_\textrm{s} \approx \gamma \lambda_n |\phi_n \rangle \langle \phi_n| +  |w \rangle \langle w| +  |f \rangle \langle f|.
\end{equation}
In the case of a closed spin chain, we have that $|\phi_n\rangle$ is the superposition state $|\phi_n\rangle = |s\rangle = \frac{1}{\sqrt{n}}\sum_{j=i}^n |j\rangle$. For our case of an open spin chain, this is still a reasonable approximation – it is correct asymptotically. Hence we introduce a state
\begin{equation}
    |p\rangle = \frac{1}{\sqrt{n-2}} \sum_{j \ne w,f}^{n} |j\rangle,
\end{equation}
such that $\langle w|p \rangle = \langle f|p \rangle = 0$ and our nearly-degenerate subspace is $\{|w \rangle, |f \rangle, |p \rangle\}$. 
The superposition state can be written in this subspace as
\begin{equation}
|s\rangle = \beta |w \rangle + \beta |f\rangle + \sqrt{1-2\beta^2} |p\rangle,
\end{equation}
where $\beta = \frac{1}{\sqrt{n}}$. Choosing $\gamma$ such that $\gamma \lambda_n \approx 1$ gives 
\begin{equation}
H_\textrm{s} \approx \begin{pmatrix} 
\beta^2 & \beta^2 & \beta\sqrt{1-2\beta^2} \\
\beta^2 & \beta^2 & \beta\sqrt{1-2\beta^2}\\
\beta\sqrt{1-2\beta^2} & \beta\sqrt{1-2\beta^2} & -2\beta^2
\end{pmatrix} + \mathds{1}.
\end{equation}
The identity matrix is neglected because it does not effect the dynamics, only adds a global phase. We have the interesting case that $\mathrm{det}(H_\mathrm{s}) = 0$, meaning we have rotations in SO(3) about an axis $\hat{n}$~\cite{curtright_elementary_2015}. Using the main result of Ref.~\cite{curtright_elementary_2015} with $\mathrm{det}(H_\mathrm{s}) = 0$, gives
\begin{equation}
    e^{-i \sqrt{2}\beta \tilde{H}_\textrm{s} t} = \mathds{1} - i \sin(\sqrt{2}\beta t) \tilde{H}_\textrm{s} + (\cos(\sqrt{2}\beta t) -1 )\tilde{H}_\textrm{s}^2,
\end{equation}
where $t$ is the time for the unitary evolution, $\sqrt{2}\beta$ is due to the normalisation $\mathrm{Tr}[\tilde{H}_{\textrm{s}}^2] = 2$, with $H_\textrm{s} = \sqrt{2}\beta \tilde{H}_{\textrm{s}}$. 

For state transfer, we want to find the time $t$ that maximises the fidelity, $F(t) = |\langle f |e^{-i H_\textrm{s} t} |w\rangle|^2$, which gives
\begin{equation}
  F(t) = \frac{\beta^2}{2} \sin^2(\sqrt{2}\beta t) + \sin^2(\tfrac{\beta t}{\sqrt{2}}).
\end{equation}
Therefore, as $n \rightarrow \infty$, $\beta=\frac{1}{\sqrt{n}} \rightarrow 0$, and we only have to consider the second term, $F \sim \sin^2(t / \sqrt{2n})$, giving state transfer at $t = \pi \sqrt{n/2}$.

The fidelity of the protocol can be approximated as the fidelity of a reverse spatial search – starting in the marked site and evolving to the superposition state – followed by a normal spatial search to the transfer site. In this case the fidelity can be analytically approximated using techniques for optimal spatial search~\cite{chakraborty_optimality_2020, lewis_optimal_2021}. 

The fidelity of the unitary evolution of this version of the protocol is
\begin{align}
    F &= \vert \langle f| e^{-i H_\mathrm{m}(f) t_2} e^{-i H_\mathrm{m}(w) t_1} | w \rangle \vert^2\\
    &= \vert \langle f| e^{-i H_\mathrm{m}(f) t_2} \sum_a |\phi_a\rangle \langle \phi_a| e^{-i H_\mathrm{m} (w) t_1} | w \rangle \vert^2
\end{align}
where $H_m(i) = \gamma H + |i\rangle\langle i|$ and $|\phi_a\rangle$ are the eigenstates of $H$. The largest eigenvalue of $H$ is, using the same approximation as before, $|\phi_n\rangle = |s\rangle$, the superposition state. Therefore, to first order
\begin{align}
    F &\approx \langle f| e^{-i H_\mathrm{m}(f) t_2} |s\rangle \langle s| e^{-i H_\mathrm{m} (w) t_1} | w \rangle \\
    &= F_{\textrm{search}}^2,
\end{align}
where $F_{\textrm{search}}$ is the fidelity of spatial search. This fidelity is reached with $t_1 = t_2$ equal to the time for optimal spatial search and gives a total time for the quantum state transfer protocol in this case as $2 T_{\textrm{search}}$ – which is actually a factor of $\sqrt{2}$ slower than the protocol we are using. The fidelity for optimal spatial search in long-range interacting systems was found in Ref.~\cite{lewis_optimal_2021}, where unit asymptotic fidelity is found for $\alpha<1$.

\section{\label{sec:transfer_simulations}State transfer simulations}
The state transfer protocol is simulated by the addition of two local fields that mark two sites, the initial site $w$ and the final site $f$. For the following results, the initial site and final site are the ends of the chain. We assume that these local fields do not alter the couplings because the local fields used are far from the spin-phonon coupling. This is achieved by a site selective AC stark shift far from the motional mode. 

Bayesian optimisation around the analytical values~\cite{lewis_optimal_2021} is used to find the optimal local fields to apply to the chain for highest fidelity quantum state transfer. The analytical approximations for the fidelity are not accurate for this low number of ions.

\begin{figure}
    \centering
    \includegraphics[scale=0.41]{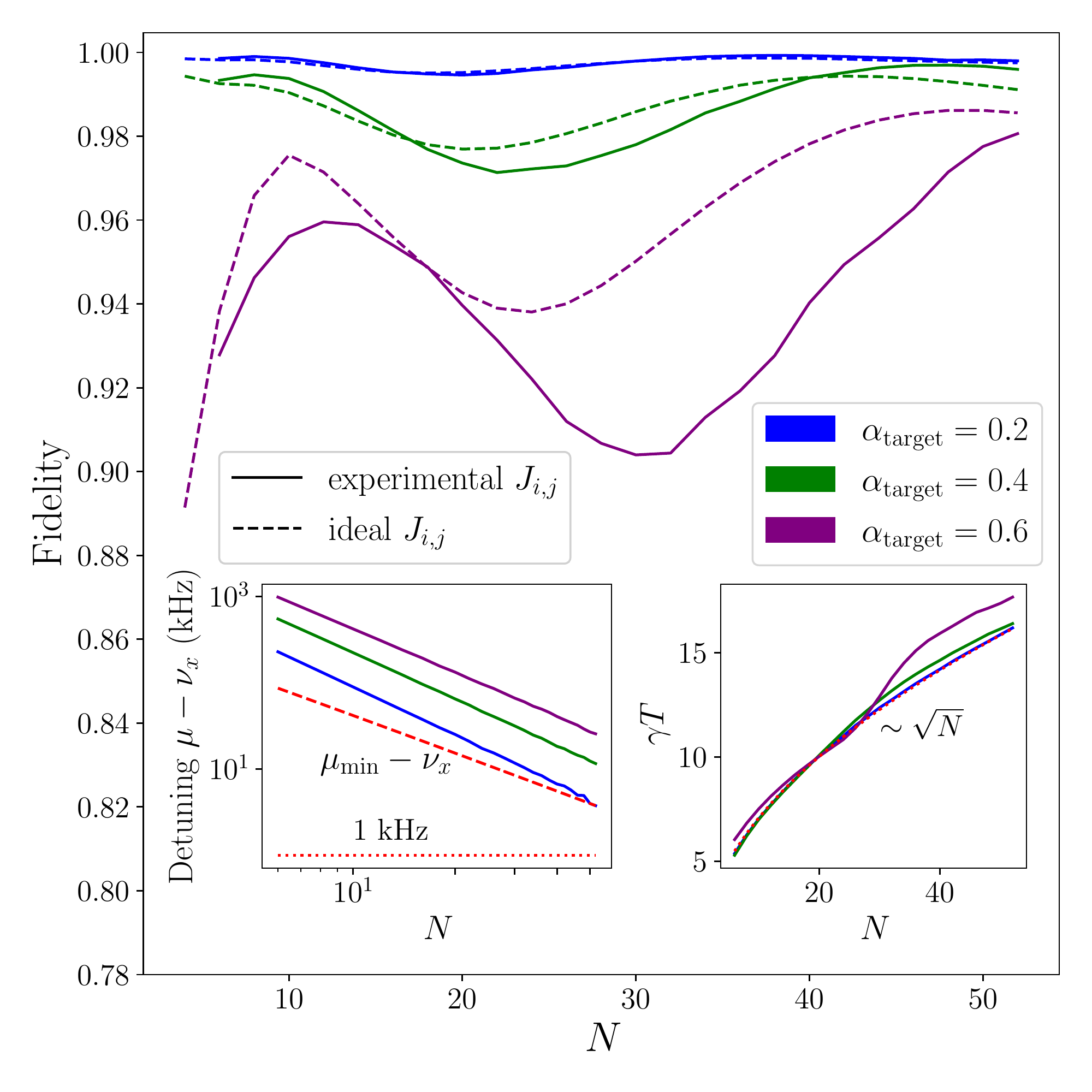}
    \caption{Fidelity of quantum state transfer protocol with experimental couplings (solid lines) $J_{i,j}$, calculated with Eq.~\eqref{eq:coupling_calculation} for $\mu$ (left inset) such that $\alpha$ is close to $\alpha_\mathrm{target}$, is compared with idealised couplings (dashed lines) $J_{i,j} \sim 1/r^\alpha$ for a chain up to $N=52$ ions. Left inset also shows the constraint in Eq.~\eqref{eq:detuning_condition}, where we have defined  $\mu_{\mathrm{min}} = 3 \Omega \eta_{m,m} + \omega_m$. Right inset shows scaled times $\tilde{T} = \gamma T$ for always-on protocol state transfer against $N$. A fit for $\sim \sqrt{N}$ is plotted (dotted red).}
    \label{fig:transfer_simulation_fideilties}
\end{figure}

In Fig.~\ref{fig:transfer_simulation_fideilties}, we use the detuning $\mu$ that results in an $\alpha$ close to the target $\alpha$ for up to a 52 ion chain. We plot the fidelity for the couplings that are experimentally motivated and for the idealised couplings $1/r^\alpha$ for $\alpha_\textrm{target}=0.2,0.4$. The results show we maintain above $\sim 0.97$ fidelity, even with the experimental couplings, and a fidelity of greater than 0.99 is possible with lower $\alpha$.

For low $N$ ions, high fidelity transfer is possible. However, the left inset in Fig.~\ref{fig:transfer_simulation_fideilties} shows that the minimum $\mu$, as defined in Section~\ref{sec:expt_design}, and therefore minimum $\alpha$, obtainable means $\alpha > 0.1$ for $N > 10$ and $\alpha > 0.2$ for $N>50$. As the number of ions increases we therefore reach the regime where $\alpha$ must be greater than 0.5, and this protocol becomes faster than direct transfer. However, even for short chains, low $\alpha$ still generates significant coherent phonons that require stroboscopic measurement, as can be seen in Fig.~\ref{fig:fidelity_xy_models}(b) for $N=10$ ions, and as $N$ increases the phonon generation magnitude $E(t)$ increases, see Appendix~\ref{sec:n_ions=8} for $N=8$ ions. In this respect, it is therefore desirable for a higher $\alpha$ to be used even for short ion chains.

The strength of the local fields for the state transfer protocol and optimal spatial search is $\gamma$, determined by the strength of the interactions between the ions. In order to meaningfully compare how the time of the protocol scales with number of ions $N$, the local fields applied must be the same. A larger $\gamma$ means stronger interactions between the ions and therefore a faster protocol time $T$. We therefore use a scaled time $\tilde{T} = \gamma T$, see right inset of Fig~\ref{fig:transfer_simulation_fideilties}, and find that $ \tilde{T} \sim \sqrt{N}$. Thus, these ion trap systems would be able to demonstrate a quantum advantage over classical algorithms for spatial search.

\section{\label{sec:noise}Noise}
\begin{figure}
    \centering
    \includegraphics[scale=0.42]{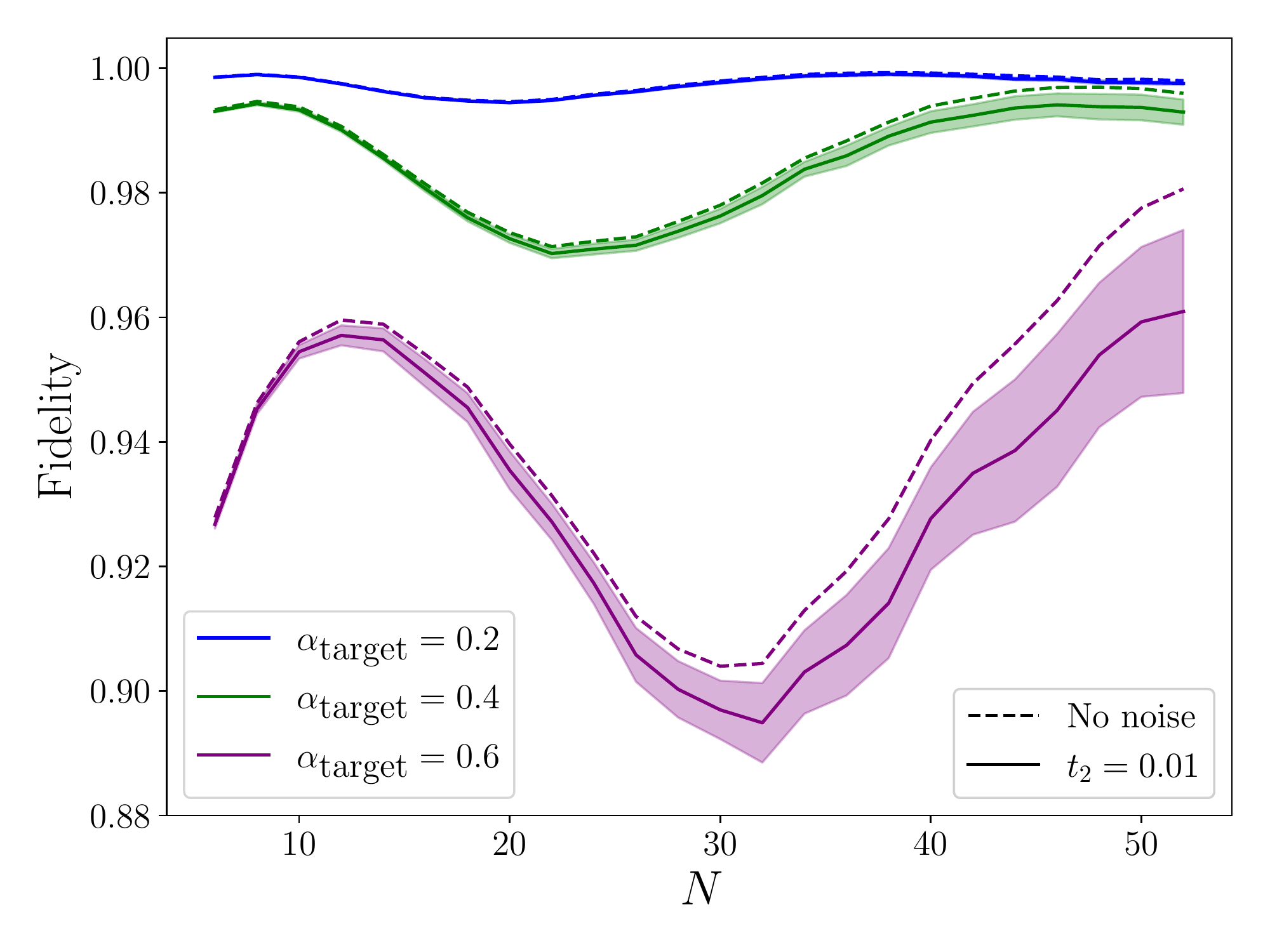}
    \caption{Dephasing time $t_2 = 10$~\si{ms} is used to compare the fidelity of quantum state transfer to the noiseless fidelity for increasing number of ions $N$. The upper and lower limits of the shaded regions are the fidelity plus or minus a standard deviation. For $\alpha_{\mathrm{target}}=0.2, 0.4, 0.6$, the experimental $\alpha$ can be achieved by optimising the detuning $\mu$. 500 samples have been used for averaging.}
    \label{fig:t2=0.01_fidelities}
\end{figure}

Dephasing noise reduces fidelity of state transfer, particularly as the number of ions $N$ increases because the time for the state transfer increases. We model dephasing noise by applying random local fields along the computational axis to every ion. We assume the random local fields do not vary on the timescale of the experiment and are normally distributed. The random local field is therefore sampled from a Gaussian distribution with variance $1/t_2$ and mean 0. The system is evolved for the full evolution. Then a new set of random fields is chosen and the system fully evolved again. Increasing the number of samples gives a better picture of the average noisy evolution. A dephasing time of $t_2 = 10$~\si{ms} has been assumed for all the ions in the presence of the Raman laser fields~\cite{monroe_programmable_2021}. The results for target $\alpha$ of 0.2 and 0.4 are shown in Fig.~\ref{fig:t2=0.01_fidelities}.

The dephasing does not have a significant effect on the transfer of a $|1\rangle$, however, it would have a larger effect on an arbitrary qubit state $|\psi\rangle = \left(|0\rangle + |1\rangle \right) / \sqrt{2}$, where the $|0\rangle$ state does not experience the same phase variation. We have assumed the noise does not vary on the timescale of the protocol. Thus, we can also give the $|0\rangle$ state the same phase by encoding the qubits as $|0\rangle \rightarrow |01\rangle$ and $|1\rangle \rightarrow |10\rangle$, and performing the state transfer for each half of the encoding – one qubit is transferred at a time, with the other not interacting with the chain. We can do this by using an additional ion at the beginning and end of the chain. The encoding can be performed with a control-NOT gate, where the control is the $|\psi\rangle$ state, followed by a NOT gate. This encoding would give approximately the same dephasing to the $|0\rangle$ and $|1\rangle$ logical states. 

\section{\label{sec:discussion}Discussion}
We have shown how an appropriate renormalisation mitigates the coherent leakage errors and enables the implementation of an optimal spatial quantum search with quadratic speedup by using the long-range interactions possible in an ion trap. This type of quantum search could potentially be an important subroutine for general algorithms in an ion trap. Of course, such algorithmic applications need further investigation, but even the demonstration of quantum speedup of spatial search in an ion trap would be interesting. A related important problem is the communication of information between distant qubits. Based on our results, we have reported a scheme which achieves a quantum state transfer between an arbitrary pair of qubits in $O(\sqrt{N})$ time if long-range interactions are available.

In Section~\ref{sec:coherent_phonons}, for two-phonon modes, the local fields can be tuned such that the effective local field terms $h_j$ of Eq.~\eqref{eq:single_site_calculation} are cancelled – Fig.~\ref{fig:fidelity_xy_models}(b) uses local fields, although the dephasing introduced when there are no initial phonons is almost imperceivable in the timescales we consider. Correcting for the local field terms is only possible if the initial phonon number is known. In general, this is not the case. However, we can use an ancillary ion to detect the phonon number as in Ref.~\cite{an_experimental_2015}. This can be measured as a preselection with an additional experimental cost: if the phonon number is 0, we perform the transfer protocol or spatial search; if the phonon number is greater than 0, we do not.

Our results correspond to a protocol for state transfer in the long-range interaction setting. The potential advantages of the protocol are twofold: first, it provides faster transfer than the direct interaction in the regime $\alpha > 0.5$, where the long-range interaction strength is characterised by the power-law decay $\sim r^{-\alpha}$; second, the protocol uses a time-independent Hamiltonian, which makes it straightforward to implement and minimises the noise that may enter the system due to required control. As we have shown, as ion chains become longer it becomes increasingly difficult to maintain the low $\alpha$ regime ($\alpha < 0.5$), and thus this protocol is a faster quantum bus for relatively long ion-chain data buses. In fact this protocol can in theory provide a transfer time that scales as $\mathcal{O}(\sqrt{N})$ with the number of ions $N$ even for $1 < \alpha <1.5$~\cite{lewis_optimal_2021}, although the fidelity of transfer is reduced.

More generally, we found that coherent phonon generation causes an effective reduction in the interaction strengths, which are given by the effective Hamiltonian of this XY model. Furthermore, we propose a scheme to find the precise factor of the reduction by fitting a leakage operator $\Vert \mathcal{E}(t)\Vert$ to the simulated phonon occupation number. This effect applies to higher excitation subspaces and when more than one phonon mode is considered, as demonstrated in Fig.~\ref{fig:fidelity_xy_models}. The effect only becomes significant for low $\alpha$, and allows lower $\alpha$ XY models to be obtained without prohibitive model error. Reaching low $\alpha$ means the time for long-range gates can be reduced, and, as we have shown in Fig.~\ref{fig:transfer_simulation_fideilties}, as the length of the spin chain $N$ increases low $\alpha$ becomes increasingly difficult. With stroboscopic measurement, these results can go some way to overcoming that obstacle.  

\section*{Acknowledgements}
DL acknowledges support from the EPSRC Centre for Doctoral Training in Delivering Quantum Technologies, grant ref. EP/S021582/1. 
LB is supported by the
U.S. Department of Energy, Office of Science, National
Quantum Information Science Research Centers, Superconducting Quantum Materials and Systems Center
(SQMS) under the contract No. DE-AC02-07CH11359.
RI and YHT acknowledge financial support from Canada First Research Excellence Fund (CFREF). RI is also supported by Natural Sciences and Engineering Research Council of Canada’s Discovery (RGPIN-2018-05250) program and Institute for Quantum Computing. SB acknowledges the support of EPSRC grants EP/R029075/1 and EP/S000267/1.

\onecolumngrid
\appendix
\section{\label{sec:alpha_definition}Definition for strength of power-law decay}
In this section, we motivate why we have defined $\alpha$ as the $1/r^\alpha$ fit of coupling strengths from one end of the chain to all other sites. Label this definition of $\alpha$ as definition A. An alternative definition, definition B, would be to include all couplings in the calculation of $\alpha$. Here, we show that for short chains, this may lead to overestimating the interaction strength for more distant interactions. It is not clear which definition is most accurate. However, we provide an argument for why we have chosen definition A.

\begin{figure}[h]
    \centering
    \includegraphics[scale=0.43]{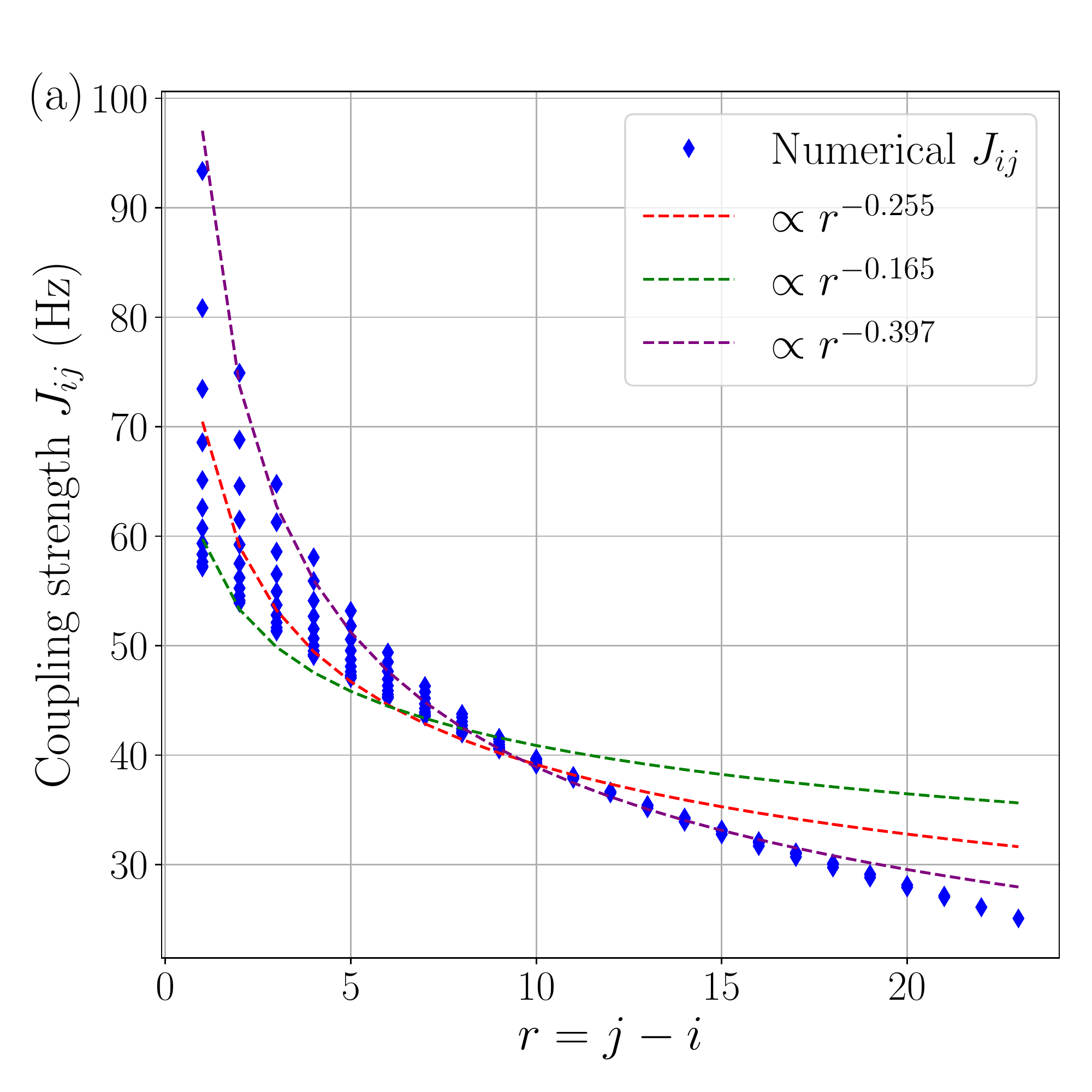}
    \includegraphics[scale=0.43]{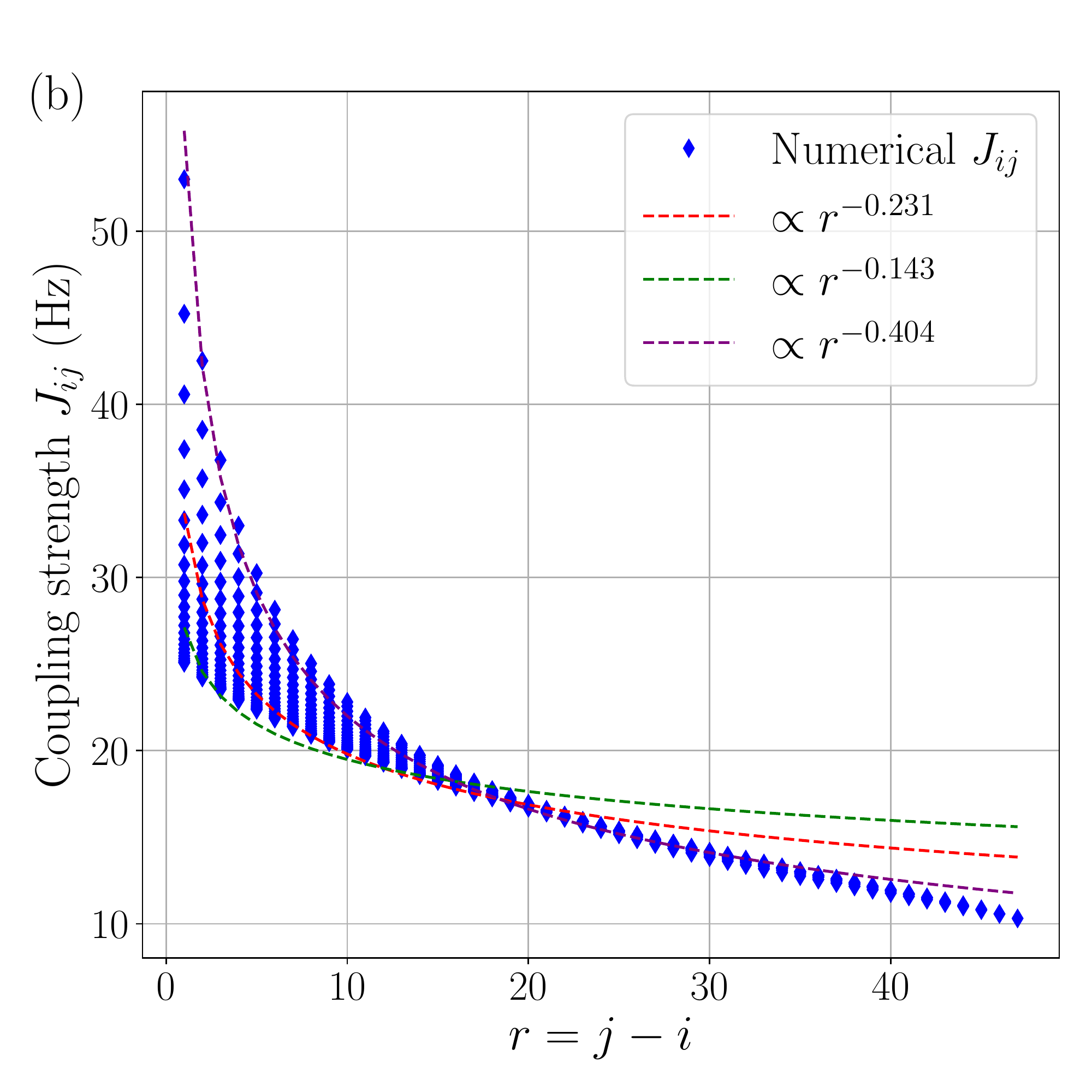}
    \caption{Coupling strength of every interaction against the difference in the ion's position in the chain for (a) 24 ion chain and (b) 48 ion chain. Three best fit curves are plotted: taking into account all interactions (red), the interactions from the centre ion (green), the interactions from the end ion (purple) – the definition of $\alpha$ used in these results.}
    \label{fig:couplings_strengths_ion_positions}
\end{figure}

In Fig.~\ref{fig:couplings_strengths_ion_positions}, the two definitions are compared for two different ion chain lengths. Firstly, definition A and definition B give significantly different $\alpha$ values. Definition A generally overestimates the interaction strengths at short range but is almost accurate for longer range interactions. Definition B is reasonable for the shorter range interactions but overestimates the long-range interactions. As a heuristic definition, we can think of higher $\alpha$ meaning less well-connected and weaker interactions. Motivated by this proposition, definition A essentially provides a lower bound, because it overestimates at the start and is accurate at the end. In the same way, definition B would also provide something like a lower bound – it is accurate at the start but overestimates the interaction strength at the end. With these heuristics in mind, and the fact we are not concerned with the coefficient of the scaling, we use definition A; definition A gives a higher $\alpha$ than definition $B$, so definition A is compatible with B giving a lower bound, but definition B is incompatible with A being a lower bound.

\begin{figure}[h]
    \centering
    \includegraphics[scale=0.43]{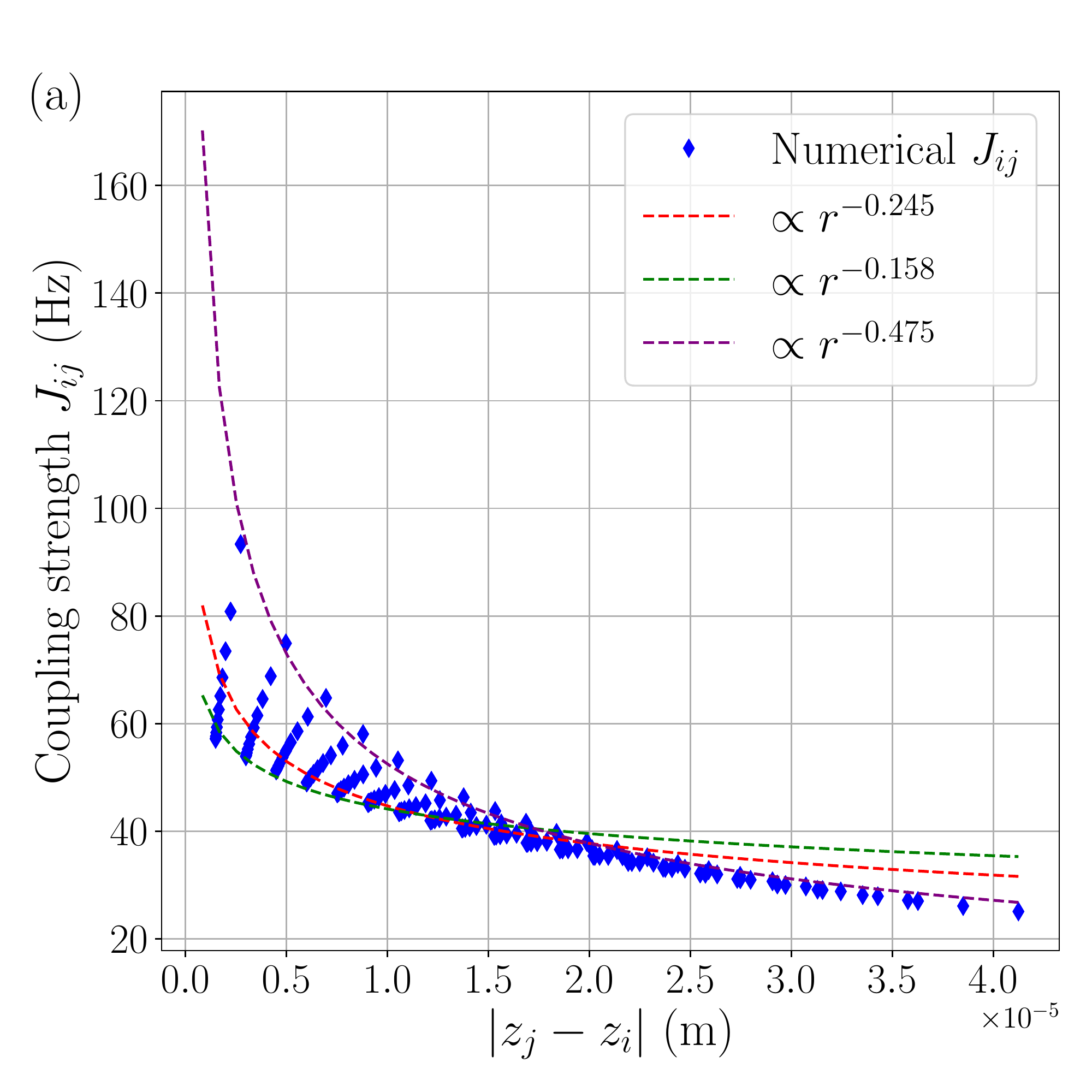}
    \includegraphics[scale=0.43]{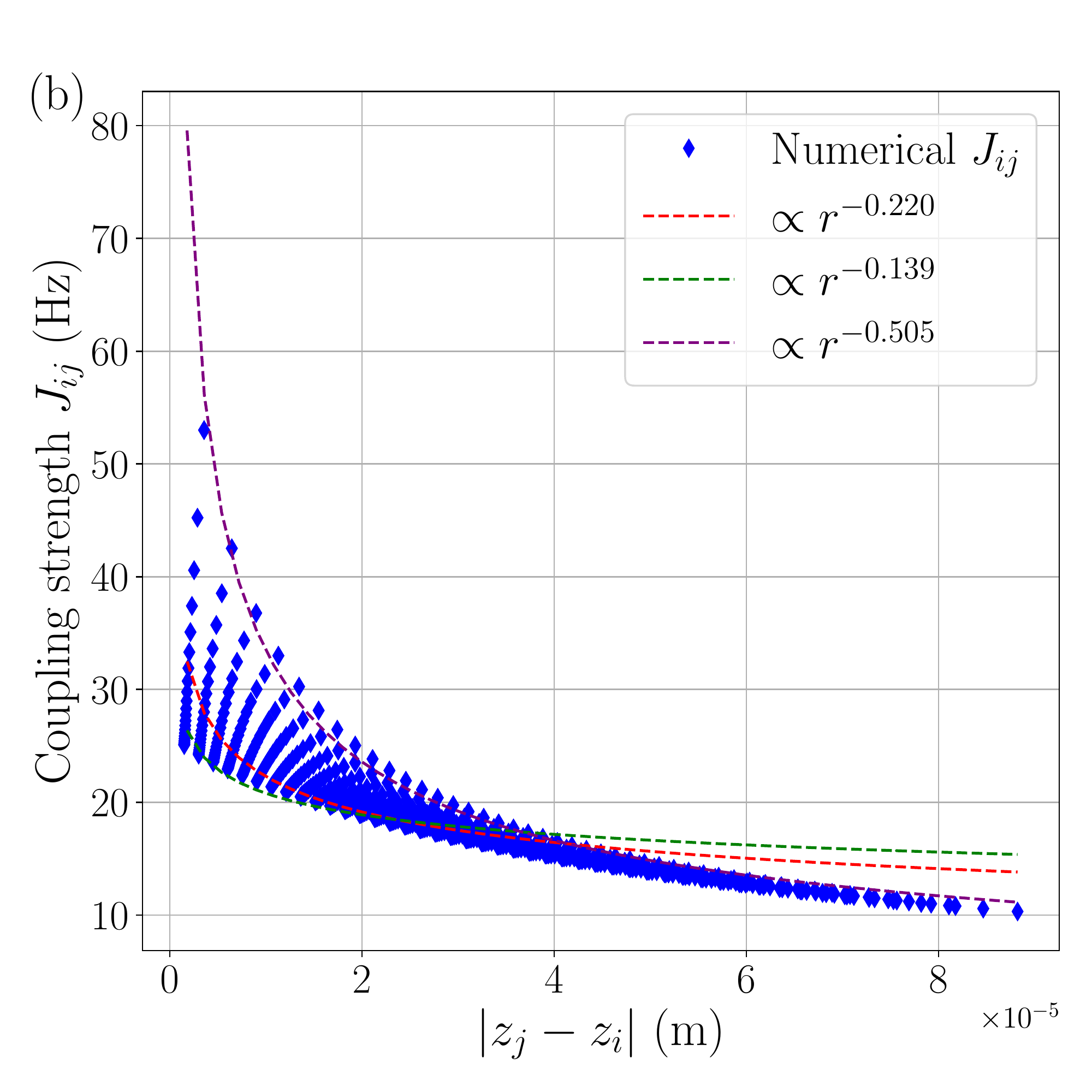}
    \caption{Coupling strength of every interaction against the difference in the ion's axial position for (a) 24 ion chain and (b) 48 ion chain. Three best fit curves: taking into account all interactions (red), the interactions from the centre ion (green), the interactions from the end ion (purple).}
    \label{fig:couplings_strengths_ion_z_positions}
\end{figure}
Rather than using position in the chain, we could also look to define $\alpha$ by using the axial position in the chain. We find the results are similar and it is not clear how to best define $\alpha$, but the preceding arguments hold, see Fig.~\ref{fig:couplings_strengths_ion_z_positions}.

\section{\label{sec:derivation_interaction_hamiltonian}Derivation of interaction Hamiltonian}
The overall Hamiltonian of an ion-phonon system with $n$ ions can be separated as   
\begin{equation}
    H = H_{\textrm{ph}} + H_{\textrm{sp}} + H_{\textrm{int}},
\end{equation}
where $H_{\textrm{ph}} = \sum_{m} \omega_m a^\dagger_m a_m$, with $a^\dagger_m$ and $a_m$ being the phonon creation and annihilation operators for mode $m$, and $H_{\textrm{sp}} = \frac{\omega_0}{2}\sum_{i} \sigma_i^z$. The interaction Hamiltonian, $H_{\textrm{int}}$, of a laser field with an ion can be modelled as an interaction of a dipole with an electromagnetic wave,
\begin{equation}
    H_{\textrm{EM-dipole}} = - \lambda E_0 \left( e^{i(k x - \omega t)}\sigma^+ + e^{-i(k x - \omega t)}\sigma^- \right).
\end{equation}
For our interaction, $k = \delta k$ is the momentum of the Raman lasers; $\omega$ is the Raman frequency, $\omega = \omega_0 + \mu$; $x$ are displacements from the equilibrium position of ion $i$ due to the phonon mode $m$, giving 
\begin{equation}
    x_i = \sum_{m} b_{im}\hat{x}_m,
\end{equation}
where $b_{im}$ is the normal mode transformation matrix, with phonon operator
\begin{equation}
    \hat{x}_m = \sqrt{\frac{\hbar}{2 M \omega_m}}(a^\dagger_m + a_m),
\end{equation}
and $M$ is the mass of the ion. We define the zero-point spatial spread of mode $m$, $\xi_m^{(0)} = \sqrt{\frac{\hbar}{2 M \omega_m}}$. Thus, we have the interaction Hamiltonian 
\begin{equation}
    H_{\textrm{int}} = \sum_{i=1}^{n}\frac{\Omega}{2} \left( e^{i(\sum_{m} \delta k b_{im} \xi_m^{(0)} (a^\dagger_m + a_m) - (\omega_0 + \mu) t)}\sigma^+_i + h.c. \right),
\end{equation}
where there are $n$ ions in the chain. The Lamb-Dicke parameter is then naturally defined as $\eta_{im} = \delta k b_{im} \xi_m^{(0)}$ and describes the strength of coupling of ion $i$ to mode $m$. In the Lamb-Dicke regime, with $\omega_0 \gg \mu \gg \Omega$, we approximate $e^{i\sum_{m} \eta_{im}  (a^\dagger_m + a_m)} \approx \Pi_m (1 + i \eta_{im} (a^\dagger_m + a_m) ) \rightarrow 1 + \sum_{m} i \eta_{im} (a^\dagger_m + a_m) $, where we have only considered the single phonon terms. Similarly, we have $e^{-i\sum_{m} \eta_{im}  (a^\dagger_m + a_m)} \rightarrow 1 - \sum_{m} i \eta_{im} (a^\dagger_m + a_m) $. Together,  
\begin{align}
    H_{\textrm{int}} &= \sum_{i}\frac{\Omega}{2} \left( e^{-i (\omega_0 + \mu) t}\sigma^+_i + h.c. \right) + \sum_{i, m}\frac{\Omega}{2} \left( i \eta_{im} (a^\dagger_m + a_m) e^{- i(\omega_0 + \mu) t}\sigma^+_i + h.c. \right),
\end{align}
with overall Hamiltonian
\begin{equation}
    H = \sum_{m} \omega_m a^\dagger_m a_m + \frac{\omega_0}{2}\sum_{i} \sigma_i^z + \sum_{i}\frac{\Omega}{2} \left( e^{-i (\omega_0 + \mu) t}\sigma^+_i + h.c. \right) + \sum_{i, m}\frac{\Omega}{2} \left( i \eta_{im} (a^\dagger_m + a_m) e^{- i(\omega_0 + \mu) t}\sigma^+_i + h.c. \right).
\end{equation}
This gives $H = H_0 + H_{\textrm{int}}$, where $H_0 =  H_{\textrm{ph}} + H_{\textrm{sp}}$.
Transforming to the rotating frame of the spins gives
\begin{align}
   H_{I_1} &=  H_{\textrm{ph}} +  e^{i(\frac{\omega_0}{2}\sum_{i} \sigma_i^z ) t} H_\textrm{int} e^{-i(\frac{\omega_0}{2}\sum_{i} \sigma_i^z ) t}.
\end{align}
Using  $[\sigma^\alpha, e^{i a n_\beta \sigma^\beta} ] = - 2 \sin(a) \varepsilon_{\alpha \beta \gamma} n_\beta  \sigma^\gamma$, where $\varepsilon_{\alpha \beta \gamma}$ is the Levi-Civita tensor and $\bm{n}$ is a unit vector giving a sum of Pauli matrices, we find
\begin{align}
    e^{i \frac{\omega_0}{2} n_\delta \sigma^\delta} \sigma^\alpha e^{ - i \frac{\omega_0}{2} n_\beta \sigma^\beta} 
    &= \cos(\omega_0 t)\sigma^\alpha + \sin(\omega_0 t) \varepsilon_{\alpha \beta \gamma} n_\beta  \sigma^\gamma + (1-\cos(\omega_0 t)) n_{\alpha}n_{\beta}\sigma^\beta, \label{eq:pauli_adjoint} 
\end{align}
where $\delta_{\alpha\beta}$ is the Kronecker delta. We note $\sigma^+ = \frac{1}{2} ( \sigma^x + i\sigma^y )$, so
\begin{align}
    e^{i \frac{\omega_0}{2} \sigma^z t}\sigma^+ e^{-i \frac{\omega_0}{2} \sigma^z t} 
    &= e^{i \omega_0 t} \sigma^+,
\end{align}
and similarly 
\begin{equation}
     e^{i \frac{\omega_0}{2} \sigma^z t}\sigma^- e^{-i \frac{\omega_0}{2} \sigma^z t} = e^{-i \omega_0 t} \sigma^-.
\end{equation}
Thus, transforming to the spins' rotating frame gives  
\begin{align}
   H_{I_1} &= H_{\textrm{ph}} + \sum_{i}\frac{\Omega}{2} \left(  e^{- i \mu t} \sigma^+_i + e^{i \mu t} \sigma^-_i \right) + \sum_{i, m}\frac{\Omega}{2} i \eta_{im} (a^\dagger_m + a_m) \left(  e^{- i \mu t} \sigma^+_i - e^{i \mu t} \sigma^-_i \right).
\end{align}
We then transform to the rotating frame of the phonons,
\begin{align}
   H_{I_2} &= \sum_{i}\frac{\Omega}{2} \left(  e^{- i \mu t} \sigma^+_i + e^{i \mu t} \sigma^-_i \right) + e^{i \sum_m \omega_m a_m^\dagger a_m t}  \sum_{i, m}\frac{\Omega}{2} i \eta_{im} (a^\dagger_m + a_m) \left(  e^{- i \mu t} \sigma^+_i - e^{i \mu t} \sigma^-_i \right) e^{-i \sum_m \omega_m a_m^\dagger a_m t} \\
   &= \sum_{i}\frac{\Omega}{2} \left(  e^{- i \mu t} \sigma^+_i + e^{i \mu t} \sigma^-_i \right) + \sum_{i, m}\frac{\Omega}{2} i \eta_{im} \left(e^{i \omega_m t} a^\dagger_m + e^{-i \omega_m t}a_m\right)\left(  e^{- i \mu t} \sigma^+_i - e^{i \mu t} \sigma^-_i \right) 
\end{align}
where we have used $e^{i \omega_m a_m^\dagger a_m t} a e^{- i \omega_m a_m^\dagger a_m t} = e^{-i \omega_m t} a$ and $e^{i \omega_m a_m^\dagger a_m t} a^\dagger e^{- i \omega_m a_m^\dagger a_m t} =  e^{i \omega_m t}a^\dagger$. Further applying the transformation $U = e^{i\frac{\mu}{2}\sum_i \sigma_i^z t}$ gives 
\begin{align}
    H_{I_3} &= \frac{\Omega}{2}\sum_{i}\sigma_i^x - \frac{\mu}{2}\sum_{i}\sigma_i^z + \frac{\Omega}{2}\sum_{i,m} i \eta_{i m}\left(e^{i \omega_m t} a^\dagger_m + e^{-i \omega_m t}a_m\right)\left( \sigma^+_i - \sigma^-_i \right), \\
    &= \frac{\Omega}{2}\sum_{i}\sigma_i^x - \frac{\mu}{2}\sum_{i}\sigma_i^z - \frac{\Omega}{2}\sum_{i,m} \eta_{i m}\left(e^{i \omega_m t} a^\dagger_m + e^{-i \omega_m t}a_m\right) \sigma^y_i 
\end{align}
Finally, we can transform to the rotating frame of $\frac{\Omega}{2}\sum_{i}\sigma_i^x - \frac{\mu}{2}\sum_{i}\sigma_i^z$,
\begin{align}
     H_{I_4} &= -\sum_{i,m} \frac{\Omega\eta_{i m}}{2}\left(e^{i \omega_m t} a^\dagger_m + e^{-i \omega_m t}a_m\right) e^{it(\frac{\Omega}{2}\sigma_i^x - \frac{\mu}{2}\sigma_i^z)} \sigma^y_i e^{-it(\frac{\Omega}{2}\sigma_i^x - \frac{\mu}{2}\sigma_i^z)}.
\end{align}
Therefore, using $\bm{n} = \frac{1}{\sqrt{\Omega^2 + \mu^2}}(\Omega, 0, -\mu)$ and the result of Eq.~\eqref{eq:pauli_adjoint}, we find
\begin{align}
     H_{I_4} &= -\sum_{i,m} \frac{\Omega\eta_{i m}}{2} \left(e^{i \omega_m t} a^\dagger_m + e^{-i \omega_m t}a_m\right) \left( \cos(\omega_{\textrm{eff}} t) \sigma_i^y -  \bm{a} \cdot \bm{\sigma} \sin(\omega_{\textrm{eff}} t) \right),
\end{align}
where $\omega_{\textrm{eff}} = \sqrt{\Omega^2 + \mu^2}$, $\bm{a} = (\mu,0,\Omega)/\omega_\textrm{eff}$. We then rotate our spin basis such that $\sigma^y \rightarrow \sigma^x$ and $\bm{a}\cdot \bm{\sigma} \rightarrow \sigma^y$, leading to the interaction Hamiltonian
\begin{align}
    H_I &= -\sum_{i,m} \frac{\Omega\eta_{i m}}{2} \left(e^{i \omega_m t} a^\dagger_m + e^{-i \omega_m t}a_m\right) \left( \cos(\omega_{\textrm{eff}} t) \sigma_i^x - \sin(\omega_{\textrm{eff}} t)\sigma_i^y \right) \label{eq:H_I} \\
    &= -\sum_{i,m} \frac{\Omega\eta_{i m}}{2} \left(e^{i \omega_m t} a^\dagger_m + e^{-i \omega_m t}a_m\right) \left( e^{i\omega_{\textrm{eff}} t} \sigma_i^+ + e^{-i\omega_{\textrm{eff}} t}\sigma_i^- \right) \\
    &= -\sum_{i,m} \frac{\Omega\eta_{i m}}{2} \left(e^{-i (\omega_{\textrm{eff}}+\omega_m) t} a_m \sigma_i^- + e^{i(\omega_{\textrm{eff}} -\omega_m) t}a_m \sigma_i^+ + h.c.\right). \label{eq:H_I_2}
\end{align}

\section{\label{sec:dyson_series}Effective Hamiltonian from Dyson series}
The effective dynamics of the interaction Hamiltonian can be derived by investigating the first and second order terms of the Dyson series. The Dyson series is a solution to the Schrödinger equation with a time-dependent Hamiltonian in the form 
\begin{equation}
    U(t) = \mathcal{T} \left[ e^{-i\int^{t}_0 d\tau H_I(\tau)}\right],
\end{equation}
where $\mathcal{T}$ is the time-ordering operator, and we have taken the initial time to be $t_0=0$. The lowest order terms of the power series of the exponential can be written
\begin{equation}
    U(t) = \mathds{1} - i \int_0^t d\tau_1 H_I(\tau_1) - \int_0^t d\tau_1 \int_0^{\tau_1} d\tau_2 H_I(\tau_1) H_I(\tau_2) + \dots
\end{equation}
Evaluating the first non-trivial term, $U_1(t) = - i \int_0^t d\tau_1 H_I(\tau_1)$, using $H_I$ from Eq.~\eqref{eq:H_I_2}, gives 
\begin{align}
     U_1(t) &= i \int_0^t d\tau_1 \sum_{i,m} \frac{\Omega\eta_{i m}}{2} \left(e^{-i (\omega_{\textrm{eff}}+\omega_m) \tau_1} a_m \sigma_i^- + e^{i(\omega_{\textrm{eff}} -\omega_m) \tau_1} a_m \sigma_i^+ + h.c. \right) \\
     &= i\sum_{i,m}  \frac{\Omega \eta_{i m}}{2}  \left( \alpha_{m}(0,0;t) a_m \sigma_i^- + \alpha_{m}(0,1;t) a_m \sigma_i^+ + h.c. \right),
\end{align}
where 
\begin{align}
     \alpha_{m}(p,q;t) &= \int^{t}_0 d\tau_1 e^{i (f_q\omega_{\textrm{eff}}+f_p\omega_m) \tau_1} \\
     &=  \frac{i\left(1-e^{i(f_q\omega_{\textrm{eff}}+f_p\omega_m)t}\right)}{(f_q\omega_{\textrm{eff}}+f_p\omega_m)},
\end{align}
and we have defined a simple function for convenience $f_k = (-1)^{k+1}$. It is clear that 
\begin{equation}
    \alpha_{m}(p,q;t)^* = \alpha_{m}(p+1,q+1;t).
\end{equation} 
The second order term is 
\begin{multline}
     U_2(t) = - \int_0^t d\tau_1 \sum_{i,m, j, l} \frac{\Omega^2 \eta_{i m}\eta_{j,l}}{4} \left( e^{-i (\omega_{\textrm{eff}}+\omega_l) \tau_1} a_l \sigma_j^- + e^{i(\omega_{\textrm{eff}} -\omega_l) \tau_1} a_l \sigma_j^+ + h.c. \right) \\ \times \left( \alpha_{m}(0,0;\tau_1) a_m \sigma_i^- + \alpha_{m}(0,1;\tau_1) a_m \sigma_i^+ + h.c. \right),
\end{multline}
this gives 
\begin{multline}
    U_2(t) = - \sum_{i,m, j, l}  \frac{\Omega^2 \eta_{i m}\eta_{j,l}}{4} \big( \beta_{l m}(0,0,0,0;t) a_l a_m \sigma_j^- \sigma_i^- + \beta_{l m}(0,0,0,1;t)  a_l a_m \sigma_j^- \sigma_i^+ \\ + \beta_{l m}(0,0,1,0;t) a_l a_m^\dagger \sigma_j^- \sigma_i^- + \beta_{l m}(0,1,0,0;t)  a_l a_m \sigma_j^+ \sigma_i^-\\
    + \beta_{l m}(1,0,0,0;t) a_m^\dagger a_l \sigma_j^- \sigma_i^- + \beta_{l m}(0,0,1,1;t)  a_l a_m^\dagger \sigma_j^- \sigma_i^+ \\
    + \beta_{l m}(0,1,0,1;t) a_l a_m \sigma_j^+ \sigma_i^+ + \beta_{l m}(1,0,0,1;t)  a_l^\dagger a_m \sigma_j^- \sigma_i^+ + h.c. \big) ,
\end{multline}
where
\begin{align}
    \beta_{l m}(r,s,p,q;t) &= \int_0^t d\tau_1 \alpha_m(p,q;\tau_1) e^{i(f_s\omega_{\textrm{eff}} + f_r\omega_l)\tau_1} \\
    &= \frac{1}{f_q\omega_{\textrm{eff}} + f_p\omega_m} \left( \frac{1-e^{i((f_q+f_s)\omega_{\textrm{eff}} + f_p \omega_m + f_r \omega_l)t}}{(f_q+f_s)\omega_{\textrm{eff}} + f_p \omega_m + f_r \omega_l} - \frac{1-e^{i(f_s\omega_{\textrm{eff}} + f_r \omega_l)t}}{f_s\omega_{\textrm{eff}} + f_r\omega_l}  \right),
\end{align}
and 
\begin{equation}
    \beta_{l m}(r,s,p,q;t)^* = \beta_{l m}(r+1,s+1,p+1,q+1;t).
\end{equation}
We then only consider the most significant terms of the $U_1(t)$ expression, essentially using the rotating wave approximation, where the denominator $\omega_\textrm{eff} + \omega_m$ suppresses the $a\sigma^-$ and $a^\dagger \sigma^+$ terms. We therefore have 
\begin{equation}
    \label{eq:U_1_RWA}
    \tilde{U}_1(t) = i \sum_{i,m} \frac{\Omega \eta_{i m}}{2} \left(\alpha_{m}(0,1;t) a_m \sigma_i^+ + \alpha_{m}(1,0;t) a_m^\dagger \sigma_i^- \right).
\end{equation}
These terms oscillate such that they are bounded and for each phonon mode $m$ the terms go to zero at multiples of time $t = \frac{2\pi}{\omega_\textrm{eff} - \omega_m}$. As the difference between $\omega_\textrm{eff}$ and the phonon mode $\omega_m$ increases these phonon excitations become less relevant~\cite{wall_boson-mediated_2017}. It is terms that scale with $t$ that contribute most to the dynamics.
In the same way, many terms in $U_2(t)$ are suppressed and some terms are bounded due to the oscillations. Ignoring these terms gives
\begin{multline}
    \tilde{U}_2(t) = - \sum_{i,m, j, l} \frac{\Omega^2 \eta_{i m}\eta_{j,l}}{4} \Big( \beta_{l m}(1,0,0,1;t)  a_l^\dagger a_m \sigma_j^- \sigma_i^+ + \beta_{l m}(0,1,1,0;t) a_l a_m^\dagger \sigma_j^+ \sigma_i^- \\
    + \beta_{l m}(1,1,0,0;t)  a^\dagger_l a_m \sigma_j^+ \sigma_i^- + \beta_{l m}(0,0,1,1;t)  a_l a_m^\dagger \sigma_j^- \sigma_i^+ \Big).
\end{multline}
We then consider just the secular terms, which are highest order and occur when $m=l$; for example
\begin{equation}
    \beta_{m m}(1,0,0,1;t) = \frac{-it}{\omega_\textrm{eff} - \omega_m} + \frac{1-e^{i(\omega_\textrm{eff} - \omega_m)t}}{(\omega_\textrm{eff} - \omega_m)^2}.
\end{equation}
This gives
\begin{multline}
    \tilde{U}_2(t) \approx - \sum_{i,j,m}  \frac{\Omega^2 \eta_{i m}\eta_{j m}}{4}\bigg(  \frac{-it}{\omega_\textrm{eff} - \omega_m}  a_m^\dagger a_m \sigma_j^- \sigma_i^+ +  \frac{it}{\omega_\textrm{eff} - \omega_m}  a_m a_m^\dagger \sigma_j^+ \sigma_i^- \\
    +  \frac{it}{\omega_\textrm{eff} + \omega_m}  a^\dagger_m a_m \sigma_j^+ \sigma_i^- + \frac{-it}{\omega_\textrm{eff} + \omega_m}  a_m a_m^\dagger \sigma_j^- \sigma_i^+ \bigg).
\end{multline}
We then use the bosonic commutation relation $[a_m, a_l^\dagger] = \delta_{ml}$,
\begin{multline}
    \tilde{U}_2(t) \approx - i t \sum_{i< j,m}   \frac{\Omega^2 \eta_{i m}\eta_{j m}}{4} \left(\frac{2\omega_m}{\omega_\textrm{eff}^2 - \omega_m^2}(\sigma_j^+ \sigma_i^- + \sigma_j^- \sigma_i^+) \right) \\
    - i t \sum_{j,m}  \frac{\Omega^2 \eta_{j m}^2  }{4} \left(\hat{n}_m \frac{2\omega_\textrm{eff}}{\omega_\textrm{eff}^2 - \omega_m^2}\left( \sigma_j^+ \sigma_j^- - \sigma_j^- \sigma_j^+ \right)+ \frac{\sigma_j^+ \sigma_j^-}{\omega_\textrm{eff} - \omega_m} - \frac{\sigma_j^- \sigma_j^+}{\omega_\textrm{eff} + \omega_m}\right),
\end{multline}
where $\hat{n}_m = a_m^\dagger a_m$. Finally, this expression can be simplified with $\sigma^+ \sigma^- = \frac{1}{2}\left(\sigma^z + \mathds{1}\right)$, $\sigma^- \sigma^+ = - \frac{1}{2}\left(\sigma^z - \mathds{1}\right)$, and $\sigma_j^+ \sigma_i^- + \sigma_j^- \sigma_i^+ = \frac{1}{2}\left(\sigma_j^x \sigma_i^x + \sigma_j^y \sigma_i^y \right)$ to give
\begin{equation}
    \tilde{U}_2(t) \approx - i t \sum_{i< j,m}   \frac{\Omega^2 \eta_{i m}\eta_{j m}\omega_m}{4(\omega_\textrm{eff}^2 - \omega_m^2)} \left( \sigma_j^x \sigma_i^x + \sigma_j^y \sigma_i^y \right)
    - i t \sum_{j,m}  \frac{\Omega^2 \eta_{j m}^2}{4} \left(\frac{\omega_\textrm{eff}}{\omega_\textrm{eff}^2 - \omega_m^2} \left(2\hat{n}_m + 1\right) \sigma_j^z - \frac{\omega_m}{\omega_\textrm{eff}^2 - \omega_m^2} \mathds{1} \right).
\end{equation}
We therefore have the effective Hamiltonian
\begin{equation}
    \label{eq:pure_xy_spin_hamiltonian_app}
    H_{XY} = \sum_{i\ne j} J_{ij}  \left( \sigma_j^x \sigma_i^x + \sigma_j^y \sigma_i^y \right) + \sum_{j} h_j \sigma_j^z,
\end{equation}
where $J_{i j}= \sum_{m} \frac{\Omega^2 \eta_{i m}\eta_{j m}\omega_m}{8(\omega_\textrm{eff}^2 - \omega_m^2)}$, $h_j = \sum_{m} \frac{\Omega^2 \eta_{j m}^2 \omega_\textrm{eff} }{4(\omega_\textrm{eff}^2 - \omega_m^2)} \left( 2n + 1 \right)$, $n$ approximates the initial phonon number, and the identity term has been dropped.
\section{\label{sec:n_ions=8}Results for 8 ions}
\begin{figure*}[h]
    \centering
    \includegraphics[scale=0.4]{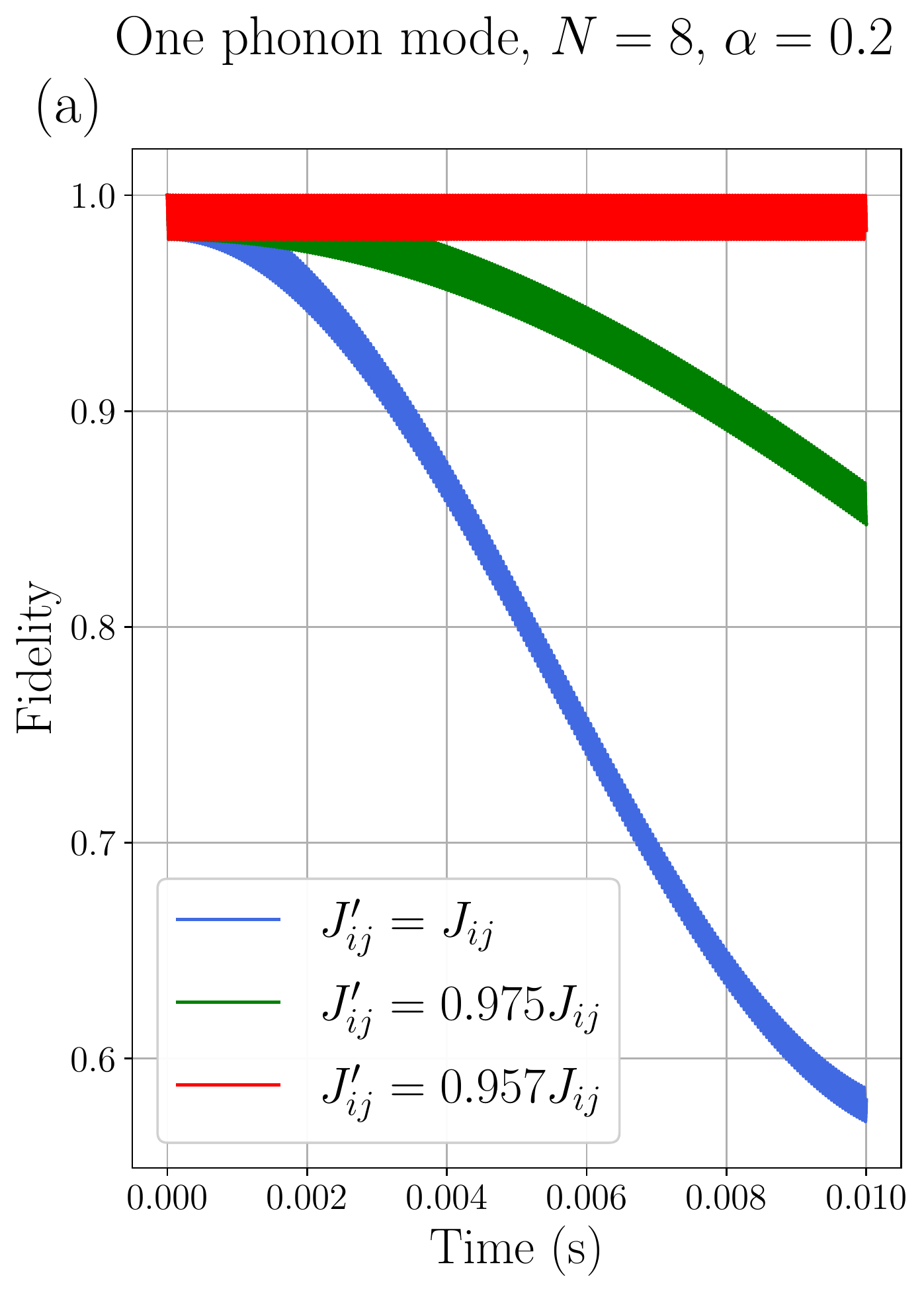}
    \includegraphics[scale=0.4]{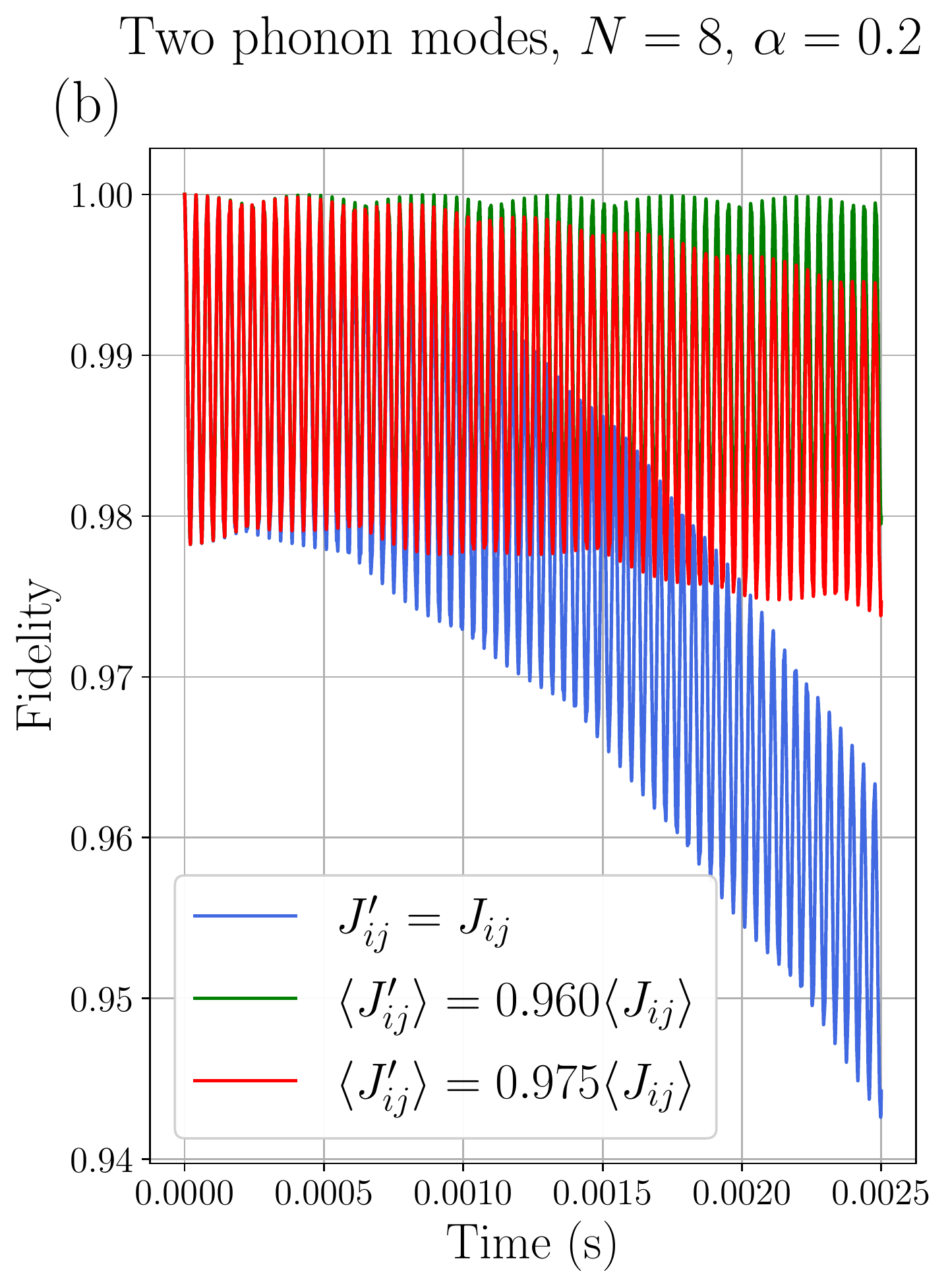}
    \includegraphics[scale=0.4]{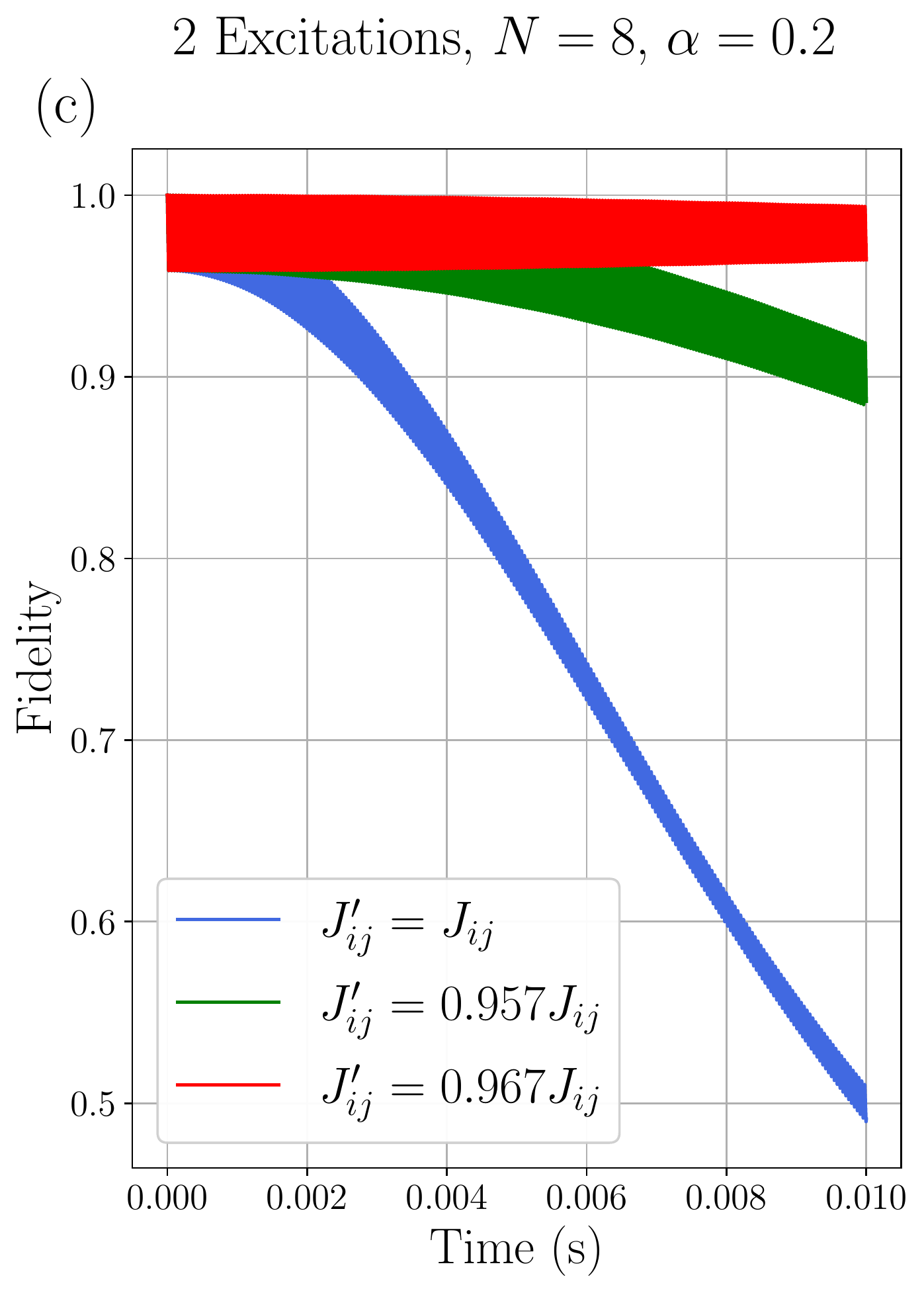}
    \caption{(a)-(b) For the initial state of no phonons and a single excited ion, the fidelity of the ion subspace is computed as for the evolution due to the full interaction Hamiltonian of Eq.~\eqref{eq:H_Interaction} for: (a) a single phonon mode with the XY model of Eq.~\eqref{eq:xy_spin_hamiltonian} for various $r$; (b) two phonon modes. The fidelity oscillates fast and regularly at frequency $\Delta_c^\prime$ with peaks and troughs separated by $\Omega^2\eta^2/\Delta_c^{\prime 2}$, as derived in Section~\ref{sec:est_coherent_phonon_generation}. The various $r$ give scaled coupling strengths, for (a) $0.957 J_{ij}$ (red), $0.975 J_{ij}$ (green) and $r=1$ simply gives $J_{ij}$ (blue); and (b) $0.960 J_{ij}$ (red), $0.975 J_{ij}$ (green), and $J_{ij}$ (blue). (c) shows the fidelity of the XY model for an initial state with two ion excitations and no phonons with $0.967 J_{ij}$ (red), $0.957 J_{ij}$ (green), and $J_{ij}$ (blue).}
    \label{fig:fidelity_xy_models_n_ions=8}
\end{figure*}

\begin{figure*}
    \centering
    \includegraphics[scale=0.4]{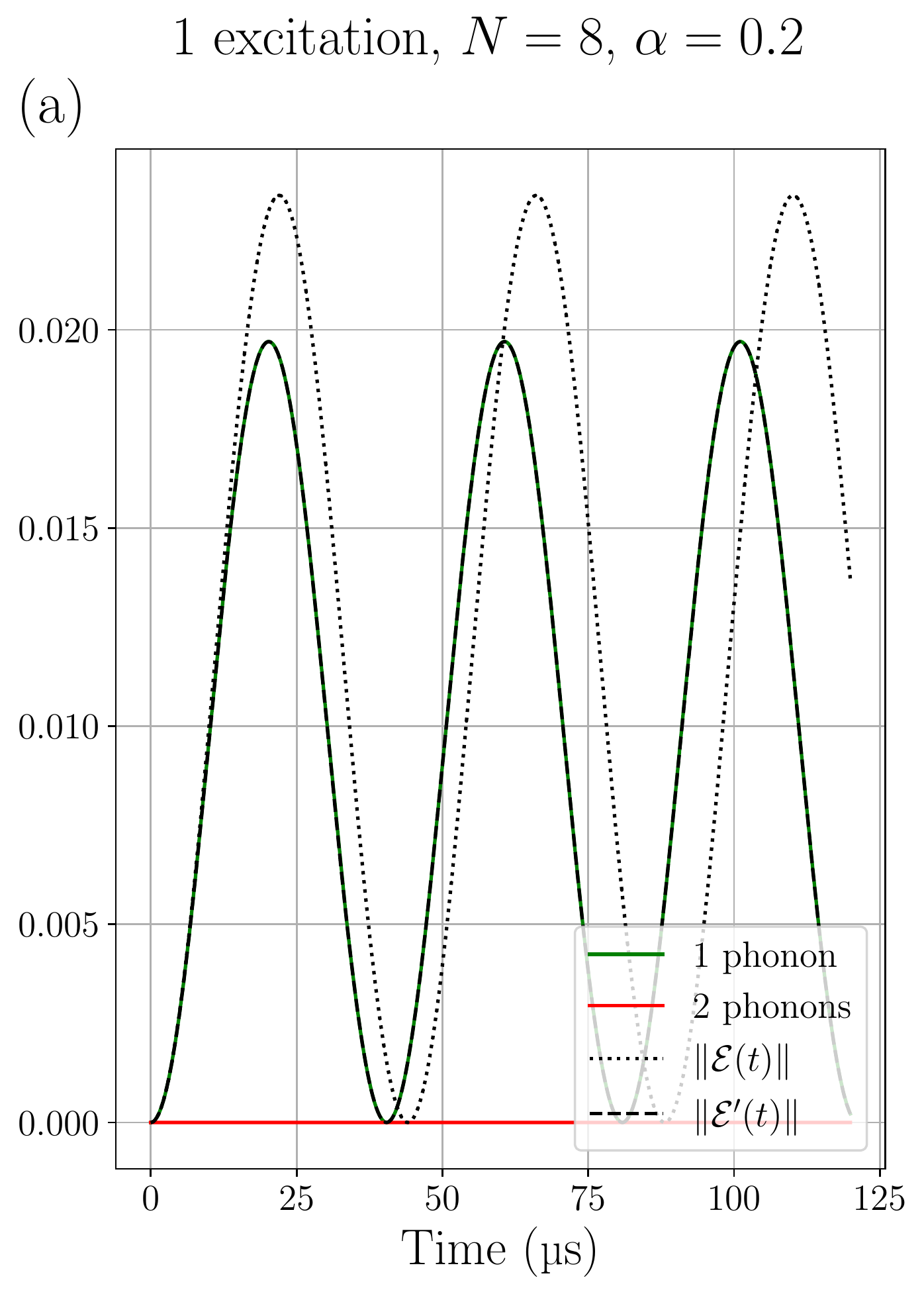}
    \includegraphics[scale=0.4]{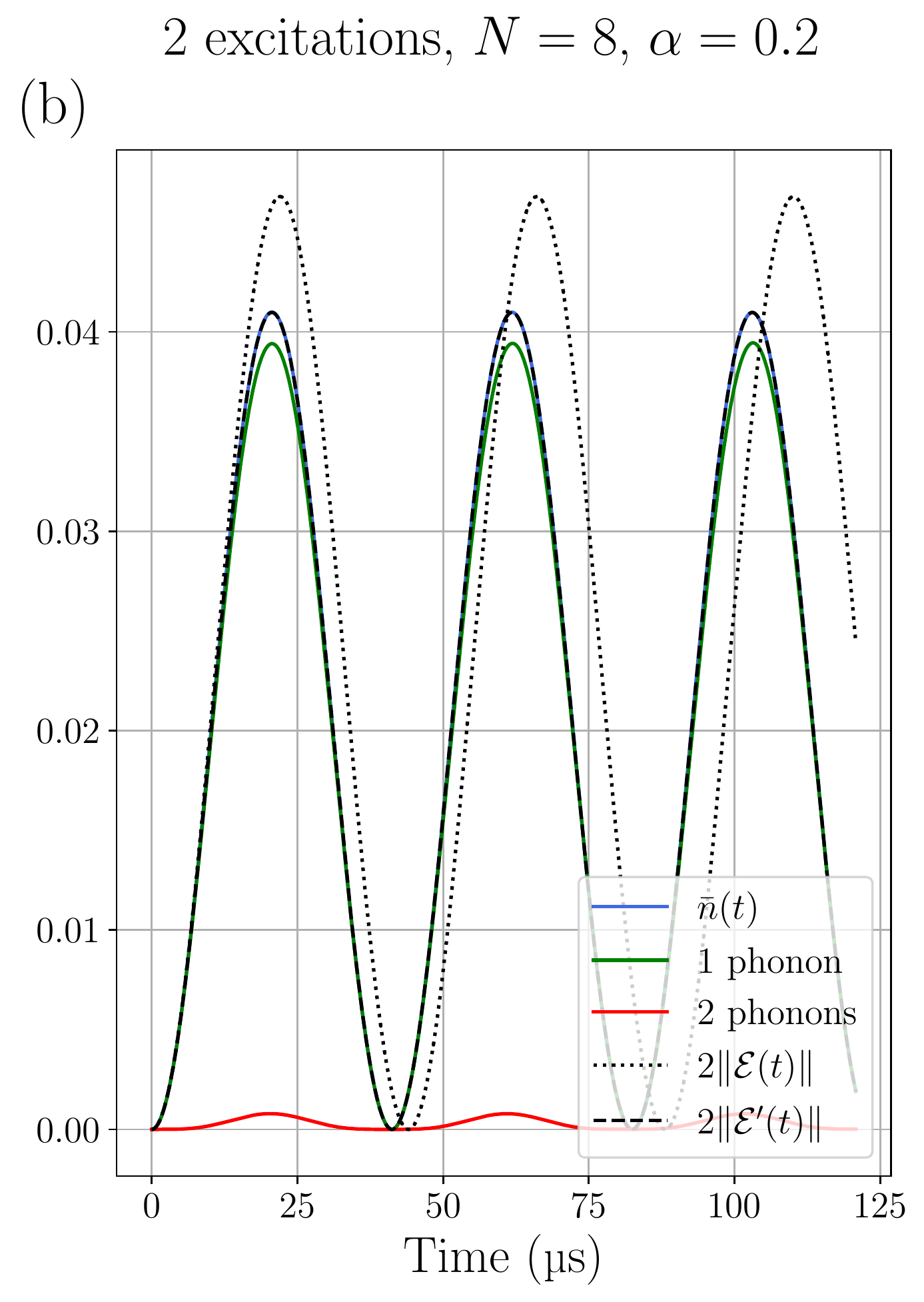}
    \includegraphics[scale=0.4]{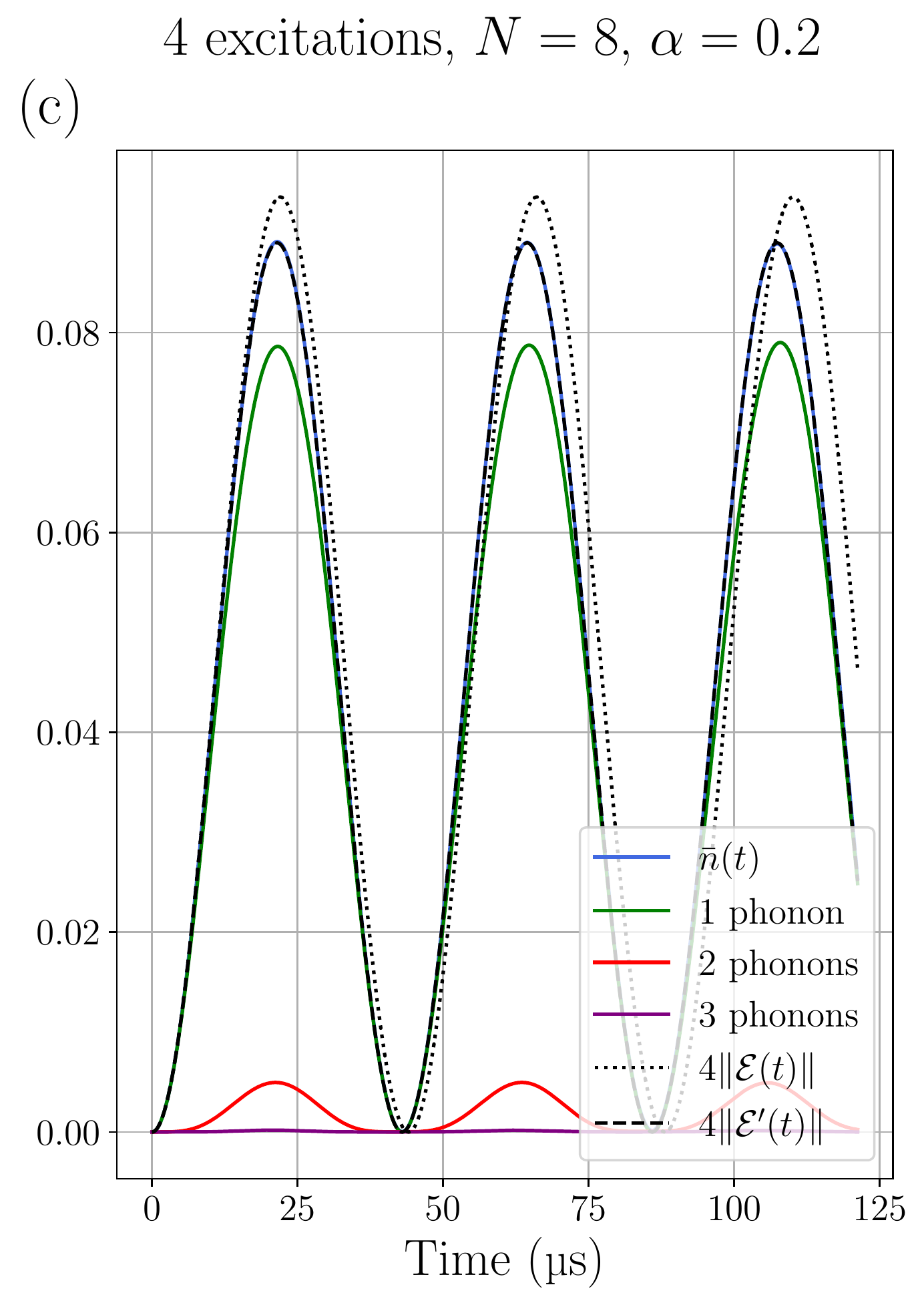}    
    \caption{Comparing the analytical $s \Vert \mathcal{E}(t) \Vert$ with the simulated phonon generation for a single phonon mode and $N=8$ ions, where $s$ is the initial number of spin excitations in the ion chain. The number of phonons is defined for $n$ phonons as $n\langle n | \textrm{Tr}_\textrm{sp} \left[ \rho(t)   \right] |n\rangle$ and $\rho(t)$ is from the evolution due to $H_I$, from Eq.~\eqref{eq:H_Interaction} with a single phonon mode. The plots show the phonon generation for a chain of various initial states: (a) shows a chain with one initial excitation, the simulated single phonon occupation number is well characterised by the leakage $\Vert \mathcal{E}^\prime(t) \Vert$ with $\omega_\textrm{eff}^\prime = 1.000338\omega_\textrm{eff}$; (b) shows a chain with two initial excitations, the simulated phonon number $\bar{n}(t)$ is well characterised by the leakage $2 \Vert \mathcal{E}^\prime (t)\Vert$ with $\omega_\textrm{eff}^\prime = 1.0002584\omega_\textrm{eff}$; (c) shows a chain with four initial excitations, the simulated phonon number $\bar{n}(t)$ is well characterised by the leakage $4 \Vert \mathcal{E}^\prime(t) \Vert$ with $\omega_\textrm{eff}^\prime = 1.000096\omega_\textrm{eff}$.}
    \label{fig:phonon_generation_higher_subspaces_n_ions=8}
\end{figure*}

The results for 8 ions are similar to that of 10 ions in the main text. This section shows that the same principles hold and the method for determining the $\omega_\textrm{eff}^\prime$ applies equally well. In fact, maintaining the same $\alpha$ requires a smaller relative shift in $\omega_\textrm{eff}$ as the number of ions decreases. Fig.~\ref{fig:fidelity_xy_models_n_ions=8} shows the fidelity of the initial state with various coupling strengths of the effective Hamiltonian. Fig.~\ref{fig:phonon_generation_higher_subspaces_n_ions=8} shoes the phonon generation for a single phonon mode and various numbers of initial excitations. 
\end{document}